# Auditing scholarly journals published in Malaysia and assessing their visibility

**A.N. Zainab [1,2], S.A. Sanni[1], N.N. Edzan [3], A.P. Koh [3]**
[1] Faculty of Computer Science & Information Technology,
University of Malaya, Kuala Lumpur, MALAYSIA
[2] Malaysian Citation Centre, Ministry of Higher Education Malaysia,
Putrajaya, MALAYSIA
[3] Library, University of Malaya, Kuala Lumpur, MALAYSIA
e-mail: zainab@um.edu.my; demolasanni@yahoo.com
edzan@um.edu.my; kohai@um.edu.my

**ABSTRACT**

*The problem with the identification of Malaysian scholarly journals lies in the lack of a current and complete listing of journals published in Malaysia. As a result, librarians are deprived of a tool that can be used for journal selection and identification of gaps in their serials collection. This study describes the audit carried out on scholarly journals, with the objectives (a) to trace and characterized scholarly journal titles published in Malaysia, and (b) to determine their visibility in international and national indexing databases. A total of 464 titles were traced and their yearly trends, publisher and publishing characteristics, bibliometrics and indexation in national, international and subject-based indexes were described.*

## INTRODUCTION AND RELATED LITERATURE

Numerous articles have been published emphasizing on the importance of scholarly journals in the dissemination and use of scientific knowledge. Related to this, especially in the early 1970s, were studies on the communication behaviour of researchers, revealing how it took longer for the social science and humanities authors to get their articles published compared to those in the physical and medical sciences (Garvey, Lin and Nelson 1970; 1971). Publishing in high impact journals therefore becomes the priority for scientists. However, the playing ground is in reality uneven. A larger majority of scholarly journals are published in the more developed countries and publish more articles by authors from developed countries. Tompkin, Ko and Donovan (2001) studied surgical journals published in the UK and US and revealed that when a journal becomes more international, there would be a drop in the percentage of articles from the country of publication and an increase in the percentage of foreign articles. In both UK and US-based journals, the numbers of contribution from authors from developing countries were low although an increase was indicated. In an editorial, Momen (2004) admitted that authors from developing countries do find difficulty in getting their articles published in high impact journals because editors like him often look for articles that have global appeal, while realizing that for these authors, global focused articles have little application in their respective countries. Their needs are therefore different or rather the reverse as they need to disseminate and gain access to literature that is relevant to their local situation, solving local problems. As a result, papers from developing countries find their way into their national journals and librarians as well as information professionals need to know what is being published. Authors also need to know who has published what and in which journals. Sadly, the bibliographic control of scholarly journals is often not done systematically in most developing countries. A number of studies have shown that Third World scholars, especially those in the science, technology and medical (STM) disciplines, still prefer to publish in foreign journals because of the problems that plague local scholarly journals (Lim 1975; Altbach 1987; Gopinathan 1992).

Twelve years ago, Normah (1999) declared in an IFLA (International Federation of Library Association) conference that there is a dearth in writings about the development of Malaysian serials as well as in controlling them bibliographically. Normah reported a total of 284 titles recorded through the issuance of International Standard Serial Number (ISSN) as reported by the National Library of Malaysia (PNM, 1996). Reports listing Malaysian scholarly journals are scattered. Md Sidin (1997) appended a list of 214 titles in a paper presented at a conference, out of which 59.3% were in the arts, humanities and the social sciences (AHSS) and 40.7% were in the science, technology and medicine. Roosfa (2006) later listed over two hundred journal titles which he classified as scholarly, and observed the difficulty of tracking journals which are increasingly mushrooming from the newly developed universities and university colleges in Malaysia. The health of a country's research activities is reflected in the publication activity of its researchers. This takes the form of its researchers' publication in journals outside the country as well as their contributions to its own national journals. As such, it is important that a country should at least have an authoritative list of all scholarly journals that are published funded through local funding which are channeled through universities, government agencies and research institutions. This paper reports on the journal audit carried out to determine the status of scholarly journals published in Malaysia, both those that have ceased or are still current.

## OBJECTIVES AND METHOD

The objective of the journal audit is to (a) to produce a master list of scholarly journals published in Malaysia; and (b) to assess the presence of these journals in universal and national indexing databases. In the Malaysian context, scholarly journals loosely refer to titles that are refereed and are research based.

There is no single current listing of Malaysian scholarly journals and trailing it requires using any existing complete listings and checking items listed against other listings located and online catalogues of universities especially the University of Malaya and the National Library of Malaysia. Checking catalogues of the former is expected to yield better results since the serials collection is expected to be more mature and complete. The former is also the oldest university in Malaysia that boasts to hold the largest collection of books and serials compared to other university libraries. The latter is the national legal repository and is expected to hold most, if not all, titles published in Malaysia. However, it is expected that not all Malaysian journal publishers are aware of their legal obligations and most often do not deposit copies of their journals with the National Library but do so at their university or institutional libraries. As a result, the complete run of journal titles are more often found in university libraries. To be able to bibliographically control and manage the listing, a small electronic solution was created and embedded in an abstracting and indexing system called *MyAIS* (*Malaysian Abstracting and Indexing System*) (http://myais.fsktm.um.edu.my). Among the menu options provided in the system is "Malaysian Journals". This is a dynamic listing, which allows instant additions, editing and deletion of titles. Only members of the audit team are allowed entry into this module and members collaboratively contribute and meet every three weeks to view titles in which the "scholarliness" is held in doubt and to iron out problems raised. The advantage of the virtual team is that a member can immediately identify duplicate or incomplete entries and could respond immediately. For each journal title, information included are the title, publisher, ISSN, frequency, year, volume and number of the first issue, whether it is print, hybrid or online, the URL address, and the indexation status of the titles in universal or national indexes. Various checklists and online catalogues are checked, information extracted and entered into the electronic listing. The sources used as checklists to identify the journals are:

- Md Sidin Ahmad Ishak, 1997, Penerbitan jurnal ilmiah di Malaysia, In: *Penerbitan Jurnal Ilmiah di Malaysia*, Kuala Lumpur, Penerbit Universiti Malaya, pp. 1-26.
- Roosfa Hashim, 2006, Penerbitan jurnal ilmiah di Malaysia: perkembangan mutakhir. *Regional Conference on Scholarly Journal Publishing*. Organized by UKM Press, 23-35 March, 2006, Pan Pacific Hotel, Kuala Lumpur, 17 p.
- *Indeks majalah Malaysia = Malaysian periodical index*, Kuala Lumpur: National Library of Malaysia, 1973 – 1990.
- *MyAIS* (*Malaysian Abstracting and Indexing System*). Available at http://myais.fsktm.um.edu.my.
- *Universiti Sains Malaysia: Jurnals and publications*. Available at http://www.usm.my/bi/main.asp?tag=penerbitan.
- Online catalogues of the University of Malaya Library and the National Library of Malaysia.
- Online catalogues of Malaysian public university libraries.

In this audit, all scholarly journal titles published in Malaysia from the earliest possible years until April 2009 are trailed. The process of listing used the following practices:

- Journal titles that change will be listed under both the old and new titles.
- Duplicates are removed.
- Journals with similar titles are made unique by including the institution, which publishes it.
- Journals that may have ceased publication are included in the list. This is especially necessary in the fields of social sciences and humanities where the useful year-life of articles is found to be longer (Zainab and Goi 1997).
- Journals which are published temporarily by a local university or college for a foreign based organization or societies, are not considered as Malaysian journals and are excluded from the list.

Visibility of the journals are determined through the indexation status in *Malaysian indexing and abstracting system* (*MyAIS*), discipline-based universal databases, and citation databases such as the *Web of Science,* and Scopus as well as universal directories such as the *Directory of Open access Journals* (*DOAJ)*.

## FINDINGS

**Publication Trends of Malaysian Scholarly Journals**
A total of 464 scholarly journal titles published in Malaysia were identified. The complete list of titles is indicated in Appendix 1. Figures 1 and 2 show the publication trends by broad year bands and yearly details for those published in 2000 onwards. The results show a continual increase, which peaks in the 2000s. This indicates a healthy publishing activity of scholarly journals. A detailed view of publishing activity in the 2000s shows higher activity before 2006, which subsequently plateaus. This may be due to the Ministry of Higher Education's emphasis on academics publishing in journals which triggered increased publishing of academic journals amongst the universities in Malaysia.

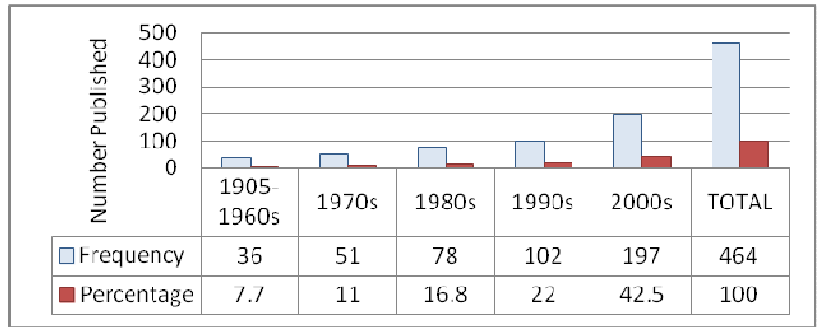

Figure 1: Total Journals Published in Broad Year Bands

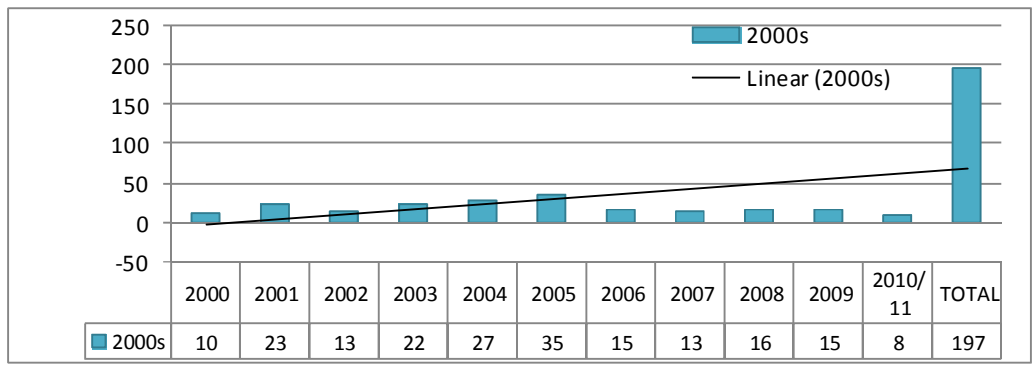

Figure 2: Trends of Publications in the 2000s

**Publishers of Malaysian Scholarly Journal**

Universities are the main publishers of scholarly journals (257, 55.5%), followed by the professional or scholarly associations (104, 22.4%) and government and private agencies (103, 22.1%) (Table 1). The older universities and colleges are more likely to be active publishers (Table 2). The total number of titles published by each university maybe exaggerated because of the inclusion of titles that may have ceased. However, we feel that the inclusion of older titles is necessary especially in fields such as history, geography, geology, cultural and ethnological studies as older findings may still be relevant and continued to be used as well as it is important for libraries in their effort to bibliographically control Malaysian serial titles.

Table 1: Publishers of Scholarly Journals in Malaysia

| Publishers | No. of Journals | Percentage |
|---|---|---|
| Government agencies | 96 | 20.6 |
| Societies, Associations | 104 | 22.4 |
| Universities, colleges | 257 | 55.5 |
| Private, companies | 7 | 1.5 |
| TOTAL | 464 | 100.0 |

Table 2: Universities/Colleges as Publishers

| Universitie/Colleges | No. of Journals | Percentage | Year Established |
|---|---|---|---|
| Universiti Malaya | 63 | 24.8 | 1962 |
| Universiti Kebangsaan Malaysia | 38 | 14.7 | 1970 |
| Universiti Teknologi Mara | 33 | 12.8 | 1999 |
| Universiti Teknologi Malaysia | 23 | 8.9 | 1973 |
| Universiti Sains Malaysia | 21 | 8.1 | 1969 |
| Universiti Islam Antarabangsa Malaysia | 13 | 5.0 | 1983 |
| Universiti Putra Malaysia | 12 | 4.6 | 1971 |
| Universiti Tun Hussein Onn Malaysia | 11 | 4.3 | 2000 |
| Universiti Perguruan Sultan Idris | 7 | 2.7 | 1997 |
| Universiti Malaysia Sabah | 7 | 2.7 | 1994 |
| Universiti Utara Malaysia | 7 | 2.7 | 1984 |
| Universiti Teknikal Malaysia Melaka | 4 | 1.6 | 2000 |
| Universiti Sains Islam Malaysia | 3 | 1.2 | 1998 |
| Universiti Malaysia Sarawak | 2 | 0.8 | 1992 |
| Universiti Malaysia Terengganu | 2 | 0.8 | 1999 |
| Universiti Malaysia Perlis | 2 | 0.8 | 2002 |
| Others (1 each) | 9 | 3.5 | |
| TOTAL | 257 | 100.0 | |

**Malaysian Journals by Broad Fields**

The fields of the arts, humanities and social sciences (AHSS) produced a larger number of journal titles (265, 57.1%) compared to those in the sciences, technology and medicine (STM) and health related fields (199, 42.9%) (Table 3). However, a number of the titles in the AHSS, may have ceased publication especially those published by the Malaysian and state level historical societies or education journals not published by the universities. This infers that perhaps journal publishers who are attached to universities in the AHSS fields were finding problems in sustaining the regularity of their issues due various reasons such as funding and poor submissions. In the STM fields, the science, medical and engineering journals are dominant with 60, 47 and 41 titles respectively (Table 4). The *Medical Journal of Malaysia* for example, publishes on average >100 articles per year and the active authors are affiliated to universities, hospitals, the Ministry of Health and private clinics (Sanni and Zainab, 2010, 2011). In Tables 3 and 4, comparisons are made with the distribution reported by a previous report (Md Sidin 1997). It is evident that since 1997 the number of journals published has increased by more than 100% in both AHSS and STM fields.

Table 3: Total Scholarly Journals Published in Malaysia by Broad Discipline

| Fields | Md Sidin (1997) | | Our results (2011) | |
|---|---|---|---|---|
| Broad fields | No. of Journals | Percentage | No. of Journals | Percentage |
| Arts, Humanities and Social Sciences (AHSS) | 127 | 59.3 | 265 | 57.1 |
| Sciences, Technology and Medicine (STM) | 87 | 40.7 | 199 | 42.9 |
| TOTAL | 214 | 100.0 | 464 | 100.0 |

Table 4: Malaysian Scholarly Journals by Fields of Study

| Fields | Md Sidin (1997)* | | Our results (2011) | |
|---|---|---|---|---|
| Fields (AHSS) | No. of Journals | Percent | No. of Journals | Percent |
| Arts, Culture, Mass Communications | 9 | 7.0 | 8 | 3.0 |
| Economics, Finance, Accounting | 22 | 17.3 | 47 | 17.7 |
| Education | 12 | 9.4 | 45 | 17.0 |
| History, Archive, Geography | 15 | 12.0 | 46 | 17.4 |
| Islam | 14 | 11.0 | 28 | 10.5 |
| Language and Literature | 5 | 4.0 | 25 | 9.4 |
| Law | 13 | 10.2 | 6 | 2.3 |
| Sociology, Library and Information Sci. | 0 | 0 | 50 | 18.9 |
| Politics, Administration, Management | 0 | 0 | 10 | 3.8 |
| General | 0 | 0 | 0 | |
| Not known | 37 | 29.1 | 0 | |
| TOTAL | 127 | 100.0 | 265 | 100.0 |
| Fields (STM) | No. of Journals | Percent | No. of Journals | Percent |
| Agriculture, Forestry, Veterinary | 11 | 12.7 | 23 | 11.6 |
| Engineering | 13 | 14.9 | 41 | 20.6 |
| Medicine and related fields | 14 | 16.1 | 47 | 23.6 |
| Others | 49 | 56.3 | 0 | 0 |
| Architecture, Building, Environment, Housing | 0 | 0 | 12 | 6.0 |
| | 0 | 0 | 0 | 0 |
| Computer Science, Information, Communication Technology | 0 | 0 | 16 | 8.0 |
| | 0 | 0 | 0 | 0 |
| Science | 0 | 0 | 60 | 30.2 |
| TOTAL | 87 | 100.0 | 199 | 100.0 |

* Source: Md Sidin (1997)

**Publishing Format of Malaysian Journals**

The majority of publishers are still publishing in print (292, 62.9%), while the rest are either, adopting the hybrid publishing policy (145, 31.3%) by maintaining both their print and electronic journals. In recent years, more of the hybrid publishers are moving towards online publishing. Very few scholarly journals are born digital (27, 5.8%). The majority of publishers are either publishing annuals (259, 55.8%) or bi-annuals (167, 36.0%)(Table 5).

Table 5: Distribution by Number of Issues Published Per Year and Publishing Format

| Number of Issues | Frequency | Percentage |
|---|---|---|
| Annual | 259 | 55.8 |
| Bi-annual | 167 | 36.0 |
| Tri-annual | 10 | 2.2 |
| Quarterly | 25 | 5.4 |
| >4 issues | 3 | 0.6 |
| TOTAL | 464 | 100.0 |
| Publishing Format | Frequency | Percentage |
| Print | 292 | 62.9 |
| Hybrid | 145 | 31.3 |
| Online | 27 | 5.8 |
| TOTAL | 464 | 100.0 |

**Visibility of Malaysian Journals in ISI *Web of Science* (*WoS*) and *Scopus***

The visibility of Malaysian scholarly journals are viewed in terms of its indexation status by national or universal indexing databases, as these databases are usually the first search points used by students and academics for literature. Out of the total 464 titles, only 105 (22.6%) are found to be indexed either by the national indexing system *MyAIS* (*Malaysian Abstracting and Indexing System*) or international indexing databases such as Thomson Reuters' indexes (*Science Citation Index, Social Sciences Citation Index, Arts and Humanities Citation Index*) or subject-based indexes (*Index Islamicus, Compendex, Chemical abstracts*, etc) (Tables 6, 7, 8).

Collectively, 51 Malaysian journals are indexed in the *Web of Science* (*WoS*) or *Scopus*. Nine (9) titles are indexed in *WoS* and 49 are indexed in Scopus. Out of the 9 titles indexed in *WoS*, 7 titles are also indexed in *Scopus*. The seven titles indexed by both databases are listed as follows.

1. *Bulletin of the Malaysian Mathematical Sciences Society* (0126-6705)
2. *Journal of Rubber Research* (1511-1768)
3. *Journal of Tropical Forest* Science (0128-1283)
4. *Malaysian Journal of Computer Science* (0127-9084)
5. *Malaysian Journal of Library and Information Science (1394-6234)*
6. *Sains Malaysiana* (0126-6603)
7. *Tropical Biomedicine* (1239-5720)

Two other titles, *Al-Shajarah (1394-6870)* and *Journal of Oil Palm Research* (1511-2780) are indexed only in the *WoS*. Out of the 49 journal titles indexed in Scopus, 42 titles are not indexed by *WoS* and listed as follows. Titles which are marked with an asterisk are no longer active.

1. *3L: Language, Linguistics, Literature* (0128-5157).
2. *ABU Technical Review* (0126-6209)
3. *ASEAN Food Journal* (0127-7324) * - superseded by title no. 9
4. *Asian Academy of Management Journal of Accounting and Finance* (1823-4992)
5. *Asia-Pacific Journal of Molecular Biology and Biotechnology* (0128-7451)
6. *Biomedical Imaging and Intervention Journal* (1823-5530)
7. *Bulletin of the Geological Society of Malaysia* (0126-6187)
8. *GEMA Online Journal of Language Studies* (1675-8021)
9. *International Food Research Journal* (1985-4668)
10. *International Journal of Asia-Pacific Studies (1823-6243)*
11. *International Journal of Business and Society* (1511-6670)
12. *International Journal of Economics and Management* (1823-836X)
13. *International Journal of Mechanical and Materials Engineering* (1823-0334)

14. *Journal of Engineering Science and Technology* (1823-4690)
15. *Journal of Science and Technology in the Tropics* (1823-5034)
16. *Journal of Sustainability Science and Management* (1823-8556)
17. *Journal of the University of Malaya Medical Centre* (1823-7339)
18. *Jurnal Ekonomi Malaysia* (0127-1962)
19. *Jurnal Pengurusan* (0127-2713)
20. *Kemanusiaan (1394-9330)*
21. *Malayan Nature Journal* (0025-1291)
22. *Malaysian Family Physician* (1985-207X)
23. *Malaysian Forester* (0302-2935)
24. *Malaysian Journal of Analytical Sciences* (1394-2506)
25. *Malaysian Journal of Consumer and Family Economics* (1511-2802)
26. *Malaysian Journal of Economic Studies* (1511-4554)
27. *Malaysian Journal of Mathematical Sciences* (1823-8343)
28. *Malaysian Journal of Medical Sciences* (1394-195X)
29. *Malaysian Journal of Medicine and Health Sciences* (1675-8544)
30. *Malaysian Journal of Microbiology* (print ISSN: 1823-8262) (online ISSN: 2231-7538)
31. *Malaysian Journal of Microscopy* (1823-7010)
32. *Malaysian Journal of Nutrition* (1394-035X)
33. *Malaysian journal of Pathology* (0126-8635)
34. *Malaysian Journal of Science* (1394-3065)
35. *Malaysian Journal of Soil Science* (1394-7990)
36. *Medical Journal of Malaysia* (0300-5283)
37. *Pertanika Journal of Science and Technology* (0128-7680)
38. *Pertanika Journal of Social Science and Humanities* (0128-7702)
39. *Pertanika Journal of Tropical Agricultural Science* (1511-3701)
40. *Tropical Life Sciences Research* (1985-8345)
41. *Sarawak Museum Journal* (0375-3050) *
42. *Warta Geologi* (0126-5539) *

The bibliometrics of the 9 journal titles indexed in *WoS* is shown in Table 6. The latest *Al-Shajarah*, published by the International Islamic University Malaysia, is indexed by the *Arts and Humanities Citation index*. Eight of the journals gained indexation status in either the *Science Citation Index* or *Social Sciences Citation Index* less than four years ago. One notable journal is *Bulletin of the Malaysian Mathematical Sciences Society*, which has shown an improvement in performance as it is ranked in Quartile 2 within the subject category of Mathematics.

Table 6: Bibliometrics of Malaysian Journals Indexed in WoS and listed in the Journal Citation Reports 2010

| No. | Title | Start Year in WoS | Impact Factor | 5-year Impact Factor | Immediacy Index | Cited Half Life | Subject category & quartile score |
|---|---|---|---|---|---|---|---|
| 1. | *Al-Shajarah* | 2010 | N/A | N/A | N/A | N/A | N/A |
| 2. | *Bulletin of the Malaysian Mathematical Sciences Soc.* | 2007 | 0.696 | N/A | 0.114 | N/A | Mathematics = Q2 |
| 3. | *J. of Oil Palm Res.* | 2007 | 0.148 | N/A | 0.000 | 7.0 | Food Sci. & Technology = Q4 |
| 4. | *J. of Rubber Res.* | 2007 | 0.154 | N/A | 0.111 | >10.0 | Polymer Science = Q4 |
| 5. | *J. of Tropical Forest Sci.* | 2003 | 0.519 | 0.428 | 0.040 | 9.4 | Forestry = Q3 |
| 6. | *Malaysian J. of Computer Science* | 2008 | 0.364 | N/A | 0.200 | N/A | Comp. Sci, Art. Intelligence = Q4, Comp. Sci, Theory & Methods = Q4 |
| 7. | *Malaysian J. of Library & information Science* | 2007 | 0.353 | N/A | 0.042 | N/A | Information Sci. & Library Sci = Q4 |
| 8. | *Sains Malaysiana* | 2008 | 0.152 | N/A | 0.06 | N/A | Multidisciplinary Sci =Q4 |
| 9. | *Tropical Biomedicine* | 2006 | 0.581 | N/A | 0.024 | 5.2 | Parasitology = Q4, Trop. Medicine= Q3 |

The bibliometrics of the 49 journal titles indexed by *Scopus* is listed in Table 7. Seven (7) titles in the list are bold and marked with asterisks to indicate that they are also indexed in *WoS*. For the eight titles, their impact scores in both *WoS* and *Scopus* are quiet different as both databases uses different sets of indicators and year frames to calculate impact scores. Bibliometric information are not reported for some titles (N/A) because either they have not fulfilled the minimum three years indexation period or their status are either inactive or have ceased publication (Table 7). The five top Malaysian Journals with highest *SCIMago Journal Ranking* (*SJR*) (*Scopus*'s quality ranking) scores are *Tropical Biomedicine* (SJR-0.053 – quartile 3: infectious diseases), *Malaysian Journal of Pathology* (*SJR*-0.047, quartile 2: Medicine, miscellaneous), *Bulletin of the Malaysian Mathematical Sciences Society* (*SJR*-0.041, quartile 2: Mathematics Misc.), *International Food Research* (*SJR*-0.040, quartile 3 0: Food sciences) and *Medical Journal of Malaysia* (*SJR*-0.040, quartile 2: Medicine, miscellaneous). The results show that the journals, which, are covered for longer number of years with reasonable *SJR* index tend to score higher on h index.

Table 7: Bibliometrics of Malaysian Journals Indexed in *Scopus* Listed in *SciMago* 2010

| | Title | Coverage | SJR Index | H-index | Subject Category & Quartile Score |
|---|---|---|---|---|---|
| 1 | *3L: Language, Linguistics, Literature* | 2008-2011 | N/A | N/A | N/A |
| 2 | *ABU Technical Review* | 1996-1999, 2001-2011 | 0.025 | 2 | Electrical & Electronic Engineering = Q4 |
| 3 | *Asian Academy of Management Journal of Accounting and Finance* | 2009-2010 | 0.026 | 1 | Accounting = Q4, Finance = Q4 |
| 4 | *Asia-Pacific Journal of Molecular Biology and Biotechnology* | 2002-2010 | 0.031 | 5 | Biochemistry = Q4, Biochemistry, Genetics & Molecular Biology= Q4, Bioengineering = Q4, Biotechnology = Q4 |
| 5 | *Biomedical Imaging and Intervention Journal* | 2005-2011 | 0.028 | 7 | Biomedical Engineering = Q4, Radiological & Ultrasound Technology = Q4, Radiology, Nuclear Medicine & Imaging = Q4 |
| 6 | *Bulletin of the Geological Society of Malaysia* | 2008-2011 | 0.025 | 1 | Earth and Planetary Science = Q4, |
| 7* | ***Bulletin of the Malaysian Mathematical Sciences Society*** | **2008-2011** | **0.041** | **4** | **Mathematics (Misc)= Q2** |
| 8 | *GEMA Online Journal of Language Studies* | 2009-2011 | 0.028 | 3 | Language & Linguistics = Q2, Linguistics & Language = Q1, Literature & Literary Theory = Q1. |
| 9 | *International Food Research Journal* | 2008-2011 | 0.040 | 5 | Food Science = Q3 |
| 10 | *International Journal of Asia-Pacific Studies* | 2011 | N/A | N/A | N/A |
| 11 | *International Journal of Business and Society* | 2009-2010 | 0.026 | 1 | Business & International Management = Q4, Economics & Econometrics = Q4 Finance = Q4, Strategy & Management = Q4 |
| 12 | *International Journal of Economics and Management* | 2006-2010 | 0.027 | 2 | Business & International Management = Q3 Economics, Econometrics & Finance (Misc) = Q3 Strategy & Management = Q3 |
| 13 | *International Journal of Mechanical and Materials Engineering* | 2007-2011 | 0.036 | 6 | Material Science (Misc) = Q3, Mechanical Engineering = Q3, Mechanics of Materials= Q3 |
| 14 | *Journal of Engineering Science and Technology* | 2009-2011 | 0.027 | 2 | Engineering (Misc) = Q3 |
| 15* | ***Journal of Rubber Research*** | **2007-2011** | **0.031** | **3** | **Organic Chemistry = Q4, Plant Science = Q3** |
| 16 | *Journal of Science and Technology in the Tropics* | 2009-2010 | 0.025 | 1 | Multidisciplinary = Q4 |
| 17 | *Journal of Sustainability Science and Management* | 2009-2010 | 0.027 | 2 | Pollution = Q4 |
| 18 | *Journal of the University of Malaya Medical Centre* | 1996-2003, 2006-2010 | 0.029 | 2 | Medicine (Misc) = Q3 |
| 19* | ***Journal of Tropical Forest Science*** | **1988 1993-2011** | **0.035** | **13** | **Agricultural & Biological Sciences(Misc) = Q3, Forestry=Q2** |
| 20 | *Jurnal Ekonomi* | 1981, 1 983, 2006-2009 | 0.026 | 0 | Business, Management & Accounting (Misc) = Q4 |
| 21 | *Jurnal Pengurusan* | 2006-2010 | 0.026 | 1 | Accounting = Q4, Business & International Management = Q4, Business, Management & |

| | | | | | Accounting (Misc) = Q4 |
|---|---|---|---|---|---|
| 22 | *Kemanusiaan* | 2008-2011 | N/A | N/A | N/A |
| 23 | *Malayan Nature Journal* | 1982-1984, 2008-2010 | 0.029 | 1 | Ecology, Evolution, Behaviour & Systematics = Q4 |
| 24 | *Malaysian Family Physician* | 2007=2010 | 0.027 | 3 | Community & Home Care = Q4, Family Practice = Q3 |
| 25 | *Malaysian Forester* | 2007-2010 | 0.026 | 2 | Forestry = Q4 |
| 26 | *Malaysian Journal of Analytical Sciences* | 2010-2011 | N/A | N/A | N/A |
| **27*** | ***Malaysian Journal of Computer Science*** | **1996-2011** | **0.030** | **5** | **Computer Science (Misc) = Q3** |
| 28 | *Malaysian Journal of Consumer and Family Economics* | 2009-2010 | 0.00 | 0 | Economics, Econometrics & Finance (Misc) = Q4 |
| 29 | *Malaysian Journal of Economic Studies* | 2007-2009 | 0.028 | 3 | Economics, Econometrics & Finance (Misc) = Q3 |
| **30*** | ***Malaysian Journal of Library & Information Science*** | **1996-2011** | **0.027** | **6** | **Library & Information Science = Q3** |
| 31 | *Malaysian Journal of Mathematical Sciences* | 2007-2011 | 0.028 | 2 | Mathematics (Misc) = Q4 |
| 32 | *Malaysian Journal of Medical Sciences* | 2003-2011 | 0.028 | 5 | Medicine (Misc) = Q3 |
| 33 | *Malaysian Journal of Medicine and Health Sciences* | 2007-2011 | 0.026 | 1 | Medicine (Misc) = Q4 |
| 34 | *Malaysian Journal of Microbiology* | 2011 | N/A | N/A | N/A |
| 35 | *Malaysian Journal of Microscopy* | 2009-2011 | 0.025 | 0 | Histology = Q4, Instrumentation = Q4 Pathology & Forensic Medicine = Q4 |
| 36 | *Malaysian Journal of Nutrition* | 2007-2010 | 0.034 | 3 | Food Science = Q3, Nutrition & Dietetics = Q3 |
| 37 | *Malaysian Journal of Pathology* | 1979-1983, 1985-2010 | 0.047 | 7 | Medicine (Misc) = Q2 |
| 38 | *Malaysian Journal of Science* | 2006-2010 | 0.030 | 4 | Multidisciplinary = Q3 |
| 39 | *Malaysian Journal of Soil Science* | 2009-2010 | 0.028 | 1 | Agronomy & Crop Science = Q4, Soil Science = Q4 |
| 40 | *Medical Journal of Malaysia* | 1963-1964, 1973-2010 | 0.040 | 13 | Medicine (Misc) = Q2 |
| 41 | *Pertanika Journal of Science and Technology* | 2010-2011 | N/A | N/A | N/A |
| 42 | *Pertanika Journal of Social Sciences and Humanities* | 2009-2012 | 0.025 | 1 | Multidisciplinary = Q4 |
| 43 | *Pertanika Journal of Tropical Agricultural Science* | 2007-2011 | 0.027 | 2 | Agronomy & Crop Science = Q4 |
| 44 | *Sains Malaysiana* | 2006-2011 | 0.031 | 4 | Multidisciplinary = Q3, |
| **45*** | ***Tropical Biomedicine*** | **2005-2011** | **0.053** | **10** | **Infectious Diseases = Q3, Parasitology = Q3,** |
| 46 | *Tropical Life Sciences Research* | 2009-2010 | 0.028 | 2 | Agricultural & Biological Sciences(Misc) = Q4, Medicine (Misc)= Q3 |
| 47 | *ASEAN Food Journal* | 2007-2008 | N/A | N/A | N/A – superseded by Int Food Res J |
| 48 | *Sarawak Museum Journal* | 1982-1996 | N/A | N/A | N/A -  Not active |
| 49 | *Warta Geologi* | 1993-1996 | N/A | N/A | N/A – Not active |

Table 8 lists 105 (22.6%) Malaysian journals and their indexation status, identified from the journal audit as in August 2011.

Table 8: Malaysian Journals Covered in International, National and Subject-Based Indexes

| Titles | Indexation Status |
|---|---|
| 1. *3L: Language, Linguistics, Literature* | Scopus, DOAJ, MyAIS, Ebsco |
| 2. *ABU Technical Review* | Scopus, Inspec |
| 3. *AFKAR Journal of Aqidah* | MyAIS |
| 4. *Akademika* | MyAIS, Index Islamicus, Sociol Abst. |
| 5. *Al-Bayan Journal of al-Quran & al-Hadith* | MyAIS |
| 6. *Al-Shajarah* | AHCI |
| 7. *Annals of Dentistry* | MyAIS |
| 8. *Archives of Orofacial Sciences* | MyAIS, WPR Index Medicus |
| 9. *Asean Food Journal* | Scopus |
| 10. *Asean J. of Teaching & Learning in Higher Education* | MyAIS, DOAJ |
| 11. *Asia Pacific J. of Molecular Biology & Biotechnology* | Scopus, MyAIS |
| 12. *Asian Acad of Management J. Accounting & Finance* | Scopus |
| 13. *Asiatic, IIUM J. of English Language and Literature* | MyAIS, Ebsco, AustLit, Bib of Asian Stud. |
| 14. *Biomedical Imaging and Intervention Journal* | Scopus, Compendex, Chem Abstr, Inspec, Doaj, Pubmed, MyAIS |
| 15. *Bulletin of the Geological Association of Malaysia* | Scopus |
| 16. *Bulletin of the Malaysian Mathematical Sciences Society* | SCI, Scopus, MathRev, MathSciNet, Zentralblatt MATH |
| 17. *e-Bangi* | ABII/INFORM, Ebsco, MyAIS, DOAJ |
| 18. *Elektrika Journal of Electrical Engineering* | MyAIS, DOAJ |
| 19. *Engineering e-Transaction* | MyAIS |
| 20. *Gema: Online Journal of Language Studies* | Scopus, Linguistics Abstr, DOAJ, MyAIS, Ebsco |
| 21. *Geografia –Malaysian Journal of Society and Space* | MyAIS, DOAJ |
| 22. *International Food Research Journal* | Scopus, Chemical Abstr, CABI, Ebsco |

| | | |
|---|---|---|
| 23. | International J Asia Pacific Studies | Scopus, Ebsco,DOAJ, Genamics, SEA Serials Index |
| 24. | International Journal of Business and Society | Scopus, Ebsco |
| 25. | International Journal of Economics and Management | Scopus |
| 26. | International Journal of Institutions and Economics | EconPapers, Scopus |
| 27. | International Journal of Mechanical and Materials Eng | Scopus, MyAIS |
| 28. | International Journal of West Asian Studies | DOAJ, Index Islamicus |
| 29. | International Medical Journal | MyAIS |
| 30. | Jebat: Malaysian Journal of History, Politics and Strategic Studies | Ebsco, MyAIS, DOAJ |
| 31. | Journal of Advanced Manufacturing Technology | MyAIS |
| 32. | Journal of Al-Tamaddun | MyAIS |
| 33. | Journal of Bioscience | MyAIS |
| 34. | Journal of Construction in Developing Countries | MyAIS |
| 35. | Journal of Engineering Science | MyAIS |
| 36. | Journal of Engineering Science and Technology | Scopus, DOAJ |
| 37. | Journal of Human Capital Development | MyAIS |
| 38. | Journal of ICT | MyAIS |
| 39. | Journal of Malaysian and Comparative Law | MyAIS |
| 40. | Journal of Mechanical Engineering and Technology | MyAIS |
| 41. | Journal of Palm Oil Research | SCI |
| 42. | Journal of Physical Science | MyAIS |
| 43. | Journal of Quality Measurement and Analysis | MyAIS |
| 44. | Journal of Rubber Research | SCI, Scopus, Cab Abstr, OVID |
| 45. | Journal of Science and Technology in the Tropics | Scopus |
| 46. | Journal of Solid State Science and Technology Letters | CAB Abstr, MyAIS |
| 47. | Journal of Sustainability Science and Management | Scopus, Chem Abstract, Zoological Records, DOAJ |
| 48. | Journal of Telecommunication Electronic and Computer Engineering | MyAIS |
| 49. | Journal of the University of Malaya Medical Centre | Scopus, MyAIS |
| 50. | Journal of Tropical Forest Science | SCI, Scopus, MyAIS |
| 51. | Jurnal Antarabangsa Teknologi Maklumat | MyAIS |
| 52. | Jurnal Ekonomi Malaysia | Scopus |
| 53. | Jurnal Fizik Malaysia | MyAIS |
| 54. | Jurnal Komunikasi, Malaysian Journal of Communication | MyAIS, DOAJ |
| 55. | Jurnal Pendidikan | MyAIS |
| 56. | Jurnal Pendidikan Malaysia | MyAIS |
| 57. | Jurnal Pengurusan (UKM) | Scopus |
| 58. | Jurnal Syariah | MyAIS, Index Islamicus |
| 59. | Jurnal Teknologi | MyAIS |
| 60. | Jurnal Usuluddin | MyAIS |
| 61. | KATHA the Official Journal | MyAIS |
| 62. | Kekal Abadi | MyAIS |
| 63. | Kemanusiaan | Scopus, MyAIS, DOAJ |
| 64. | Labuan Bulletin of Int Business & Finance | MyAis |
| 65. | Labuan e-journal of Muamalat and Society | MyAIS |
| 66. | Malayan Nature Journal | Scopus |
| 67. | Malaysian Family Physician | Scopus |
| 68. | Malaysian Forester | Scopus |
| 69. | Malaysian Journal of Analytical Sciences | Scopus, MyAIS, INIS |
| 70. | Malaysian Journal of Biochemistry and Molecular Biology | MyAIS |
| 71. | Malaysian Journal of Civil Engineering | MyAIS |
| 72. | Malaysian Journal of Community Health | Medline, IndexMedicus |
| 73. | Malaysian Journal of Computer Science | SCI, Scopus, Inspec |
| 74. | Malaysian Journal of Consumer and Family Economics | Scopus |
| 75. | Malaysian Journal of Economic Studies | Scopus, ABI/Inform, Econlit |
| 76. | Malaysian Journal of Leaning & Instruction | MyAIS |
| 77. | Malaysian Journal of Library & Information Science | SSCI, Scopus, LibLit, LISA |
| 78. | Malaysian Journal of Mathematical Science | Scopus |
| 79. | Malaysian Journal of Medical Sciences | Scopus, MyAIS |
| 80. | Malaysian Journal of Medicine and Health Sciences | Scopus |
| 81. | Malaysian Journal of Microbiology | Scopus, DOAJ, MyAIS |
| 82. | Malaysian Journal of Microscopy | Scopus |
| 83. | Malaysian Journal of Nutrition | Scopus, MyAIS |
| 84. | Malaysian Journal of Pathology | Scopus, MyAIS |
| 85. | Malaysian Journal of Pharmaceutical Sciences | MyAIS, DOAJ, CAB Abstr |
| 86. | Malaysian Journal of Psychiatry | MyAIS |
| 87. | Malaysian Journal of Science | Scopus, MyAIS |
| 88. | Malaysian Journal of Social Administration | MyAIS |
| 89. | Malaysian Journal of Soil Science | Scopus |
| 90. | Malaysian Management Journal | MyAIS |
| 91. | Malaysian Online Journal of Instruction Technology | MyAIS |
| 92. | Malaysian Orthopaedic Journal | MyAis |
| 93. | Malaysian Polymer Journal | MyAIS |
| 94. | Masalah Pendidikan | MyAIS |
| 95. | *Matematika Jurnal* | MyAIS |
| 96. | *Medical Journal of Malaysia* | Scopus, Medline, MyAIS |
| 97. | *Pertanika Journal of Trop Agric Sci* | Scopus |
| 98. | *Pertanika Journal of Sc & Tech* | Scopus |
| 99. | *Pertanika Journal of Social Science & Humanities* | Scopus |
| 100. | *Sains Malaysiana* | SCI, Scopus, ChemAbstr, Zentralblatt MATH, MyAIS |
| 101. | *Sarawak Museum Journal* | Scopus |
| 102. | *Sari Jurnal Alam dan Tamadun Melayu* | MyAIS |
| 103. | *Tropical BioMedicine* | SCI, Scopus, Medline, Pubmed |
| 104. | *Tropical Life Sciences Research* | Scopus MyAIS |
| 105. | *Warta Geologi* | Scopus |

## a) Publishers of Malaysian Journals Indexed by WoS and Scopus

The public and private universities published a total of 29 (57%) journal titles out of 51, which are indexed in *WoS* and *Scopus*. The remaining 29% (15), are produced by professional associations and the rest (7, 14%) are from government and semi-government agencies (Table 9). There are cases where a department at a University publishes for a Malaysian-based association or society and these are categorized under the respective Universities. Examples are, *Bulletin of the Malaysian Mathematical Sciences Society* is listed under Universiti Sains Malaysia who publishes it for the Society, *Malaysian Journal of Economic Studies* is placed under Universiti Malaya who publishes for the Malaysian Economics Association and *Malaysian Journal of Microscopy* is placed under Universiti Putra Malaysia who publishes it for Electron Microscopy Society Malaysia.

Table 9: Publishers of Malaysian Journals Indexed in WoS and Scopus

| | Publisher | Titles | Indexation |
|---|---|---|---|
| 1. | Academy of Family Physicians Malaysia | *Malaysian Family Physician* | Scopus |
| 2. | Academy of Science Malaysia | *Journal of Science and Technology in the Tropics* | Scopus |
| 3. | Asean Food Handling Bureau | *ASEAN Food Journal* | Scopus |
| 4. | Asian Broadcasting Union | *ABU Technical Review* | Scopus |
| 5. | Forest Research Institute Malaysia | *Malaysian Forester* | Scopus |
| 6. | Forest Research Institute Malaysia | *Journal of Tropical Forest Science* | WoS (SCI), Scopus |
| 7. | Institute of Mathematical Research Malaysia | *Malaysian Journal of Mathematical Sciences* | Scopus |
| 8. | Mal Soc for Biology & Biotechnology | *Asia-Pacific Journal of Molecular Biology and Biotechnology* | Scopus |
| 9. | Malaysian Analytical Science Society | *Malaysian Journal of Analytical Sciences* | Scopus |
| 10. | Malaysian Consumer and Family Economic Association | *Malaysian Journal of Consumer and Family Economics* | Scopus |
| 11. | Malaysian Geological Society | *Bulletin of the Geological Society of Malaysia* | Scopus |
| 12. | Malaysian Geological Society | *Warta Geologi* | Scopus |
| 13. | Malaysian Nature Society | *Malayan Nature Journal* | Scopus |
| 14. | Malaysian Nutrition Society | *Malaysian Journal of Nutrition* | Scopus |
| 15. | Malaysian Pathologist Society | *Malaysian Journal of Pathology* | Scopus |
| 16. | Malaysian Society for Microbiology | *Malaysian Journal of Microbiology* | Scopus |
| 17. | Malaysian Society for Parasitology and Tropical Medicine | *Tropical Biomedicine* | WoS (SCI), Scopus |
| 18. | Malaysian Society of Soil Science | *Malaysian Journal of Soil Science* | Scopus |
| 19. | Medical Association of Malaysia | *Medical Journal of Malaysia* | Scopus |
| 20. | Palm Oil Research Institute Malaysia | *Journal of Oil Palm Research* | WoS (SCI) |
| 21. | Rubber Research Institute Malaysia | *Journal of Rubber Research* | WoS (SCI), Scopus |
| 22. | Sarawak Museum Dept | *Sarawak Museum Journal* | Scopus |
| 23. | Taylors University College | *Journal of Engineering Science and Technology* | Scopus |
| 24. | Universiti Islam Antarabangsa | *Al-Shajarah* | WoS (AHCI) |
| 25. | Universiti Kebangsaan Malaysia | *3L: Language, Linguistics, Literature* | Scopus |
| 26. | Universiti Kebangsaan Malaysia | *GEMA Online Journal of Language Studies* | Scopus |
| 27. | Universiti Kebangsaan Malaysia | *Jurnal Pengurusan* | Scopus |
| 28. | Universiti Kebangsaan Malaysia | *Sains Malaysiana* | WoS (SCI), Scopus |
| 29. | Universiti Kebangsaan Malaysia | *Jurnal Ekonomi Malaysia* | Scopus |
| 30. | Universiti Malaya | *Biomedical Imaging and Intervention Journal* | Scopus |
| 31. | Universiti Malaya | *International Journal of Mechanical and Materials Engineering* | Scopus |
| 32. | Universiti Malaya | *Journal of the University of Malaya Medical Centre* | Scopus |
| 33. | Universiti Malaya | *Malaysian Journal of Computer Science* | WoS (SCI), Scopus |
| 34. | Universiti Malaya | *Malaysian Journal of Library & Information Science* | WoS (SSCI), Scopus |
| 35. | Universiti Malaya | *Malaysian Journal of Science* | Scopus |
| 36. | Universiti Malaya for Malaysian Economic Association | *Malaysian Journal of Economic Studies* | Scopus |
| 37. | Universiti Malaysia Sarawak | *International Journal of Business and Society* | Scopus |
| 38. | Universiti Malaysia Terengganu | *Journal of Sustainability Science and Management* | Scopus |
| 39. | Universiti Putra Malaysia | *International Food Research Journal* | Scopus |
| 40. | Universiti Putra Malaysia | *International Journal of Economics and Management* | Scopus |
| 41. | Universiti Putra Malaysia | *Malaysian Journal of Medicine and Health Sciences* | Scopus |
| 42. | Universiti Putra Malaysia | *Pertanika Journal of Science and Technology* | Scopus |
| 43. | Universiti Putra Malaysia | *Pertanika Journal of Social Science and Humanities* | Scopus |
| 44. | Universiti Putra Malaysia | *Pertanika Journal of Tropical Agricultural Science* | Scopus |
| 45. | Universiti Putra Malaysia for Electron Microscopy Society Malaysia | *Malaysian Journal of Microscopy* | Scopus |
| 46. | Universiti Sains Malaysia | *Asian Academy of Management Journal of Accounting and Finance* | Scopus |
| 47. | Universiti Sains Malaysia | *International Journal of Asia-Pacific Studies* | Scopus |
| 48. | Universiti Sains Malaysia | *Malaysian Journal of Medical Sciences* | Scopus |
| 49. | Universiti Sains Malaysia | *Kemanusiaan* | Scopus |
| 50. | Universiti Sains Malaysia | *Tropical Life Sciences Research* | Scopus |
| 51. | Universiti Sains Malaysia for the Society | *Bulletin of the Malaysian Mathematical Sciences Society* | WoS (SCI), Scopus |

### Broad Subject Categorization of Journals Indexed in WoS and Scopus

Within *WoS* and *Scopus*, the journals are assigned subject categories (Table 10). One journal title may be assigned to more than one subject category. Most of the Malaysian journals indexed in *WoS* and *Scopus* are in the Sciences category (18, 36%). Three (3) journals are categorized in the miscellaneous category under science, namely, J*ournal of Science and Technology in the Tropics* (Malaysian Academy of Science), *Malaysian Journal of Science* (Universiti Malaya), and *Sains Malaysiana* (Universiti Kebangsaan Malaysia). Three (3) titles are In the Arts and Humanities category, *3L: language, linguistics, Literature* and *Gema: Online Journal of Language Studies* both published by Universiti Kebangsaan Malaysia and *Al-Shajarah* published by the International Islamic University Malaysia.

Table 10: Broad Field Categorization of Malaysian Journals Indexed in *WoS* and *Scopus*

| Broad fields | No. of Journals | Percentage |
|---|---|---|
| Science | 18 | 35.3 |
| Medicine | 10 | 19.6 |
| Engineering | 4 | 7.8 |
| Arts, Humanities | 5 | 9.8 |
| Social Sciences | 10 | 19.6 |
| Miscellaneous (Science) | 4 | 7.8 |
| TOTAL | 51 | 99.9 |

### CONCLUSION

The audit, tracing of Malaysian scholarly journals resulted in a master list of titles that have been carefully selected. Journals that have at least a 5-year publishing run are included even though they have ceased publication. This is especially so in the fields of the arts and social sciences where older articles often remain relevant to researchers and important for libraries, which can use such listing to identify gaps in their collections. From the audit, a list of scholarly journal titles which can be dynamically updated is maintained at the *Malaysian Abstracting and Indexing System* (*MyAIS*) through it sub-menu "Malaysian journals". The master list provides libraries with a useful source for tracing gaps in their collection. Over 90% of the titles are available at the University of Malaya Library and other university libraries since, over 50% of the journals are published by universities and colleges. The list is also useful for the national indexing and citation database to identify the extent of journals that needs coverage. Such a citation system is being developed by the recently established

Malaysian Citation Centre commissioned and financed by the Ministry of Higher Education Malaysia in February 2011. The Centre is currently located at the University of Malaya. The list also provides students and researchers with a useful source to identify relevant journals to peruse or submit articles for publication.

If in the early 1990s there were concerns about the effects of scientific communication by electronic publishing on the scholarly communication process (Harnad, 1997; Rowland, 1997; Sosteric, 1996), the 2000s saw growing support for e-journal publishing (Liew, Foo, Chennupati, 2000). This is evident amongst the Malaysian journals as more publishers are adopting the hybrid publication policy in the 1990s, maintaining both the print and electronic versions of their journals. Subsequently, a number of the hybrid publishers in recent years have switched to online only publishing mode to cope with publishing costs and to close submission-to-publication gap. The first known digital electronic journal is the *Journal of Problem- based Learning* published by the Faculty of Computer Science and Information Technology, University of Malaya in 2003 but this journal has ceased publication. The next e-journal was the *Malaysian Online Journal of Instructional Technology,* which, published its first issue in 2004 and this journal has also not kept itself current as the latest online version is volume 3, no.3, 2006. A noteworthy digital born journal is the *Biomedical Imaging and Intervention Journal*, published in 2005 by the Department of Biomedical Imaging, Faculty of Medicine, University of Malaya, which is currently indexed by *Scopus, Compendex, Chemical Abstract, Inspec* and *PubMed*. This is one of the few journals that use a .org domain to publish and is successful in maintaining its issues and currency. The number of publishers who has opted to publish only online currently numbers 27, many of which were previously hybrid.

While trailing Malaysian journals, we found journals which do not conform in characteristics to qualify as Malaysian journals and have been excluded in the Malaysian Journal Master list. These are journals that are published by a department or an institution in Malaysia for foreign-based associations or societies. We identify three such titles that are in this category, *Neurology Asia* (ISSN:1823-6138), produced by the Neurology Laboratory at the Universiti Malaya Medical Centre for the Asean Neurological Association, *Asia-Pacific Journal of Public Health* (ISSN:1010-5395), produced by the Health Research and Development Unit, Faculty of Medicine, Universiti Malay for the Asia Pacific Academic Consortium for Public Health and *Asian Myrmecology* (ISSN: 1985-1944), recently produced by the Institute for Tropical Biology and Conservation, University Malaysia, Sabah for the International Network for the Study of Asian Ants. Besides this, we also found foreign publishers which have a branch in Malaysia publishing journals in the fields of computer science and engineering. We refer to journals published by CSC Journals, a publisher which has a branch in Kuala Lumpur, Melbourne and Lahore and does not qualify as a Malaysian publisher. *The Directory of Open Access Journals* (*DOAJ*) have listed six titles by this publisher under Malaysia, *International Journal of Computer Science and Security* (ISSN:1985-1553), *International Journal of Biometrics and Bioinformetics* (ISSN: 1985-2347), *International Journal of Engineering* (1985-2312), *International Journal of Image Processing* (1985-2304), *International Journal of Security* (ISSN:1985-2320) and *Signal Processing: an International Journal* (ISSN: 1985-2339).

There is also an increase in the number of journals being indexed either by *MyAIS* the Malaysian indexing and abstracting system or by international indexing databases. This indicates a growing consciousness amongst journal editorials to improve quality and increase chances of getting indexed. This is also in line with Malaysian Ministry of Education's call to improve quality and indexation status. In December 2010, the Meeting of Cabinet Ministers Malaysia approved the proposal for setting up of a Malaysian Citation Centre. This centre was unofficially set up in February 2011 and is responsible for indexing scholarly journals published in Malaysia as well as providing bibliometric reports on the performance of the Malaysian journals. This is the beginning of the proper monitoring and bibliographic control of Malaysian scholarly journals.

Malaysian Journal Master List.

1. **3L Southeast Asian Journal of English Language Studies**.
Universiti Kebangsaan Malaysia. School of Language Studies and Linguistics. *ISSN:0128-5157* (Annual. Vol.1, 1991). Hybrid. http://www.ukm.my/ppbl/3L/3LHome.html). Indexation: Scopus, DOAJ, MyAIS.

2. **ABU Technical Review.**
Asian Broadcasting Unions Communication. *ISSN:0126-6209* (6 times a year. Jan, Mar, May, Jul, Sept, Nov. 1987-) Print. Indexation: Inspec, Scopus.

3. **Accacmadia Journal.**
Institut Teknologi Mara. School of Accountancy. *ISSN:0126-9577.* (2-4 times a year. Vol.1,1975 - ). Print.

4. **AFKAR Journal of Aqidah and Islamic Thought**.
University of Malaya. Academy of Islamic Studies. *ISSN:15118819* (Annual. Vol.1, 2000 - E-version: http://bakkdev.um.edu.my/myjurnal/public/browse-journal-view.php?id=186) Hybrid. Indexation: MyAIS.

5. **Akademika Journal**.
Universiti Kebangsaan Malaysia. *ISSN:01265008.* (Triannual,Vol.1, 1972). Hybrid. http://www.ukm.my/penerbit/jademik.html ). Indexation: Index Islamicus, CSA Sociological Abstract, MyAIS.

6. **Al-Bayan Journal of Al-Quran & al-Hadith**.
University of Malaya. Academy of Islamic Studies. *ISSN:1394-3723.* (Biannual. Vol.1, 2003 -). Hybrid. http://bakkdev.um.edu.my/myjurnal/public/browse-journal-view.php?id=165). Indexation: MyAIS.

7. **Al-Jazari International Journal of Civil Engineering.**
Universiti Teknologi MARA. *ISSN:1823-7681* (Annual. 2006 -). Print.

8. **Al-Mizan : Jurnal Sekolah Pengajian Islam.**
KUSZA. *ISSN:1675-1523* (Annual. Bil. 1, 2002- ). Print.

9. **Al-Shajarah.**
International Islamic University Malaysia. *ISSN:13946870.* (Annual, Vol.1, no.1, 1996). Print. Indexation: Arts and Humanities Citation Index.

10. **Al-Sirat: Jurnal Kolej Islam Pahang Sultan Ahmad Shah .**
Kolej Islam Pahang Sultan Ahmad Shah. (Annual. Vol.1, 2001-). Print.

11. **Alam Melayu Jurnal.**
University of Malaya. Akademi Pengajian Melayu. *ISSN:0128-6927* . (Annual. Jil.1, 1993 - Jil.2, 1994-). Print. Ceased. Title changed to *Jurnal Pengajian Melayu*.

12. **Alam Sekitar Journal.**
Persatuan Perlindungan Alam Sekitar Malaysia (Malaysian Environment Protection Society). *ISSN:0126-7280.* (Quarterly. Vol.1, 1976 - ). Print.

13. **Analisis: Jurnal Universiti Utara Malaysia**.

Universiti Utara Malaysia. *ISSN:0127-8983* (Annual. Vol.1,1986 - Vol.10, 2003) Hybrid. http://ijms.uum.edu.my/index.php?option=com_phocadownload&view=section&id=2&Itemid=12.

14. **Annals of Dentistry**
University of Malaya. Faculty of Dentistry. *ISSN:0128-7532.* (Annual. Vol.1, 1994 - ). Hybrid. http://ejum.fsktm.um.edu.my/VolumeListing.aspx?JournalID=13). Indexation: MyAIS.

15. **Archives of Orofacial Sciences**
Universiti Sains Malaysia, School of Dental Sciences. *ISSN:1823-8602* . (Annual. Vol.1, 2006 - ) Hybrid. http://www.dental.usm.my/aos/) Indexation: MyAIS, Index Medicus.

16. **Asean Food Journal**.
Asean Food Handling Bureau, Kuala lumpur. *ISSN:01277324* (Triannual. Vol.1 no 1,1985-2007) . Changed title to *International Food Research Journal*. Print.Indexation: Scopus, Chemical Abstracts, Biological Abstracts, MyAIS.

17. **Asean Journal of Teaching and Learning in Higher Education.**
Universiti Kebangsaan Malaysia. *ISSN:19855826* . (Biannual, Vol1, no.1, 2009 -) Hybrid. http://www.ukm.my/jtlhe/Archives.aspx ). Indexation: MyAIS, DOAJ.

18. **Asia Pacific Journal of Educators and Education.**
Universiti Sains Malaysia. School of Educational Studies. *ISSN:0126-7272.* (Annual, Vol.25, 2010 -). Former title: *Jurnal Pendidik dan Pendidikan*. Hybrid. http://web.usm.my/education/content.asp?m=pub&n=journal).

19. **Asia Pacific Journal of Molecular Biology & Biotechnology.**
Malaysian Society for Biology & Biotechnology. *ISSN:01287451.* (Triannual. Vol.1 no.1, 1993-) Hybrid. http://www.msmbb.org.my/apjhome.htm. Indexation: Scopus, MyAIS.

20. **Asia Pacific Journal of Molecular Medicine.**
Universiti Kebangsaan Malaysia, Medical Molecular Biology Institute. *ISSN:223200326. Online* (Annual, Vol.1, 2011-). http://www.umbi.ukm.my/umbi/index.php?option=com_content&task=view&id=388&Itemid=74 )

21. **Asia Pacific Management Accounting Journal.**
Universiti Teknologi MARA. *ISSN:1675-3194* . (Annual. Vol.1, 2004-). Hybrid. http://ari.uitm.edu.my/journals/jofra/jofra-vol-4-no-1-2006.html.

22. **Asian Academy of Management Journal of Accounting and Finance.**
Universiti Sains Malaysia for Asian Academy of Management (AAM). *ISSN:18234992* . (Biannual Vol.5, 2000). Print. Indexation: Scopus, Ebsco.

23. **Asian Journal of Business and Accounting.**
University of Malaya. Faculty of Business and Accounting. *ISSN:19854-64* . (Biannual. Vol.1 no.1, 2008-). Hybrid. http://umrefjournal.um.edu.my/journal/index.php?menu=view_detail&id=51)

24. **Asian Journal of University Education.**
Universiti Teknologi MARA. *ISSN:1823-7797*. Absorbed: *Malaysian Journal of University Education*. (Biannual. Vol.1, 2005-). Hybrid. http://acrulet.net/asian_journal_of_university_education.

25. **Asiatic, IIUM Journal of English Language and Literature**.
International Islamic University Malaysia. Dept. of English Language and Literature. *ISSN:19853016.* (Annual Vol. 1, 2007 - ). Hybrid. http://asiatic.iium.edu.my/) Indexation: EBSCO, AustLit: Resource for Australian Literature, Bibliography of Asian Studies, MyAIS, Journal of Commonwealth Literature's Annual Bibliography,UK).

26. **ASM Science Journal**.
Akademi Sains Malaysia.*ISSN:1823-6782* (Biannual. Vol.1 no.1, 2007- ). Hybrid. http://www.akademisains.gov.my/index.php?option=com_content&task=view&id=102&Itemid=245.

27. **At-Tajdid**
International Islamic University Malaysia. Al-Jami`ah al-Islamiyah al-`Alamiyah, Petaling Jaya. *ISSN:1823-1926* . (Biannual. No.1, 1997-). Hybrid.http://rms.research.iium.edu.my/bookstore/Category/75-wwgooglecom.aspx)

28. **Beringin: Jurnal Akademi Seni Kebangsaan.**
Akademi Seni Kebangsaan. *ISSN:1823-6383* .(Annual. Vol.1, 2005-). Print.

29. **Biomedical Imaging and Intervention Journal.**
University of Malaya. Faculty of Medicine. Department of Biomedical Imaging. *ISSN:18235530* .(Quarterly. Vol.1, 2005-). Online. http://biij.org/ Indexation: Scopus, Embase, Compendex, Chemical Abstracts, Inspec, DOAJ, PubMed, MyAIS.

30. **Biotech Communications.**
Universiti Putra Malaysia. Faculty of Biotechnology and Biomolecular Sciences. *ISSN:1823-3279* . (Biannual. Vol.1, 2005-). Print.

31. **Borneo Research Journal (Jurnal Penyelidikan Borneo).**
University of Malaya, Dept of Southeast Asian Studies. *ISSN:19855443.* (Annual. Vo.1, 2007 – Vol.3, 2009) Print.

32. **Borneo Review : Journal of the Institute for Development Studies Sabah.**
Institute for Development Studies Sabah. *ISSN:0128-7397).* (Annual. Vol. 1 no. 1, 1990- ). Print.

33. **Borneo Science: a Journal of Science and Technology.**

148. **Journal of Mechanical Engineering**
Universiti Teknologi MARA. Research Management Institute. *ISSN:1823-5514.* (Biannual. Vol.1, 2004-). Hybrid. http://rmi.uitm.edu.my/journal-of-mechanical-engineering.html)

149. **Journal of Mechanical Engineering and Technology.**
Fakulti Kejuruteraan Mekanikal, Universiti Teknikal Malaysia Melaka *ISSN:21801053.* (Annual. Vol.1, no.1, 2009 -). Hybrid. http://jmet.utem.edu.my/index.php?option=com_frontpage&Itemid=1) Indexation: MyAIS

150. **Journal of Media and Information Warfare.**
Universiti Teknologi MARA. *ISSN:1985-563X .*(Annual. Vol.1, 2008-). Partial Hybrid. http://cmiws.uitm.edu.my/publications/journals/journal-volume-1.html)

151. **Journal of Modern Languages**
University of Malaya. Faculty of Language & Linguistics. *ISSN:1675-526X .* (Annual. Vol.1, 1980 ). Hybrid from 2009 issue. http://jml.um.edu.my/current)

152. **Journal of Muamalat and Islamic Finance Research**
Kolej Universiti Islam Malaysia. Faculty of Economics and Muamalat *ISSN:1823-075X* (Annual. Vol.1, no.1, 2004- ). Hybrid. http://ddms.usim.edu.my/handle/123456789/2151?show=full)

153. **Journal of Nuclear and Related Technologies.**
Malaysian Nuclear Society. ISSN:*1823-0180.* (Annual .Vol.1, 2001 - ). Hybrid. Online from 2004 – http://www.nuklearmalaysia.org/index.php?id=14&mnu=14).

154. **Journal of Occupational Safety & Health.**
National Institute of Occupational Safter & Health Malaysia. *ISSN: 1675-5456.* (Biannual. Vol.1, no.1, 2004 –). Hybrid. http://www.niosh.com.my/osh-journal.html.

155. **Journal of Oil Palm Research.**
Palm Oil Research Institute of Malaysia. *ISSN:1511-2780. Palmolis* (Biannual. Vol.10, no.1 1998 -). Hybrid. http://jopr.mpob.gov.my/. Indexation: Scopus.

156. **Journal of Physical Sciences.**
Universiti Sains Malaysia. School of Chemical Sciences. *ISSN: 1675-3402, Online: 2180-4230.* (Biannual. Vol,1, no.1, 1990). Online 2006-. Hybrid. http://web.usm.my/jps/index.html Indexation: MyAIS.

157. **Journal of Plant Protection in the Tropics.**
Persatuan Perlindungan Tumbuh-Tumbuhan Malaysia. *ISSN:0127-6883 .* (Biannual. Vol.1, no.1,1984 -). Print.

158. **Journal of Problem-Based Learning**.
University of Malaya. Faculty of Computer Science and Information Technology. (Annual. Vol.1, 2003 – 2005). Online. Ceased. *http://ejum.fsktm.um.edu.my/VolumeListing.aspx?JournalID=9.* Indexation status: MyAIS

159. **Journal of Quality Measurement and Analysis (JQMA)**.
Universiti Kebangsaan Malaysia. *ISSN:1823-5670* (Biannual. Vol.1, no.1, 2005 - ). Hybrid. http://www.ukm.my/~ppsmfst/jqma/index2.html. Indexation status: MyAIS

160. **Journal of Rubber Research.**
Institut Penyelidikan Getah Malaysia. *ISSN:1511-1768 .* (Quarterly. Vol.1 1929 - ) Hybrid. http://rios.lgm.gov.my/cms/fedDigiJournalDetail.jsp?searchText=&digiCon=true&selTab=digiCon&type=JRR&id=&issueYear=2010. First published as the *Journal of the Rubber Research Institute of Malaya* in 1929 and *Journal of Natural Rubber Research* in 1986. Indexation: SCI, Scopus. CAB Abstracts and OVID.

161. **Journal of Science and Mathematics Education in Southeast Asia.**
Regional Educational Centre, Science and Mathematics Education, Penang. *ISSN:0126-7663 .*(Biannual. Vol.1, no.1, 1978-). Print.

162. **Journal of Science and Technology.**
Universiti Tun Hussein Malaysia. *ISSN:2229-8460.* (Biannual. Vol.1, no.1, 2009-) Hybrid. http://penerbit.uthm.edu.my/component/content/article/1-umum-2/70-call-for-a-paper.html).

163. **Journal of Science and Technology in the Tropics.**
Akademi Sains Malaysia & Confederation of Scientific & Technological Association in Malaysia (COSTAM). *ISSN:1823-5034.* (Biannual, Vol.1 No.1, 2005-). Hybrid.Online 2009. http://www.akademisains.gov.my/index.php?option=com_content&task=view&id=330&Itemid=336. Indexation: Scopus.

164. **Journal of Solid State Science and Technology Letters.**
Malaysian Solid Science and Technology Society. *ISSN:0128-8393 ,* (Biannual. Vol.1, 1994-2008-) . Hybrid. 2005-2007 online. http://www.mass-malaysia.net/letters/lsystem/index.php?page=volume-15-no-2-2008.

165. **Journal of Sustainability Science and Management.**
Universiti Malaysia Terengganu. *ISSN: 1823-8556.* (Vol.1, no.1, 2006). Online from issues 2006. Indexation: Scopus, Zoological Records, Chemical Abstracts, DOAJ.

166. **Journal of Technical Education and Training.**
Universiti Tun Hussein Malaysia. *ISSN:22298932.* (Biannual. Vol.1, no.1, 2009-). Hybrid. http://penerbit.uthm.edu.my/ejournal/index.php?option=com_content&view=category&layout=blog&id=15&Itemid=22

167. **Journal of Techno-Social.**
Universiti Tun Hussein Malaysia. *ISSN 2229-8940* . (Biannual, Vol.1, no.1, 2009-). Hybrid. http://penerbit.uthm.edu.my/ojs/index.php/JTS.

168. **Journal of Telecommunication Electronic and Computer Engineering**
Universiti Teknikal Melaka Malaysia. ISSN: 1985-3158. (Biannual. Vol.1, no.1,

2009-). Online. http://jtec.utem.edu.my/index.php?option=com_frontpage&Itemid=1. Indexation: MyAIS.

169. **Journal of the Federated Malay States Museum.**
Museums Dept Malaysia. *ISSN:Z090832159.* (Annual. Vol.1, 1905 - 1941-). 1906-1922 online at this site. http://www.biodiversitylibrary.org/bibliography/22352

170. **Journal of the Malayan Branch of the Royal Asiatic Society.**
Malayan Branch of the Royal Asiatic Society.(Annual. Vol.1, 1923 - 1963-). Print.

171. **Journal of the Malaysian Branch of the Royal Asiatic Society.**
Malaysian Branch of the Royal Asiatic Society of Gt. Britain and Ireland. *ISSN:0126-7353.* (Annual. Vol.37, 1964). Print

172. **Journal of the Malaysian English Language Teaching Association.**
Malaysian English Language Teaching Association. *ISSN:1511-8002. 44* (Annual, 2005 – 2009-). Hybrid. *http://www.melta.org.my/modules/tinycontent/index.php?op=edit&id=*

173. **Journal of the University of Malaya Medical Centre (JUMMEC).**
University of Malaya Medical Centre. *ISSN:1823-7339.*(Annual.1, 1996). Hybrid. http://jummec.um.edu.my/ Indexation: Scopus. MyAIS.

174. **Journal of Tourism, Hospitality And Culinary Arts**
Universiti Teknologi MARA. *ISSN: 1985-8914.* (Annual. Vol.1, 2011-). Hybrid. http://www.jthca.org/journal/index.php?option=com_content&view=article&id=65&Itemid=77.

175. **Journal of Tropical Forest Science**
Forest Research Institute of Malaysia *ISSN:0128-1283.* (Quarterly. Vol.1, no.1, 1988- ). Hybrid. *E-version:* http://www.frim.gov.my/?page_id=1826. Indexation. SCI, Scopus, MyAIS.

176. **Journal of Tropical Medicinal Plants.**
Tropical Botanics. *ISSN:1511-8525.* (Biannual, Vol.1, no.1/2, 2001-). Print.

177. **Journal of Wildlife and Parks.**
Jabatan Perlindungan Hidupan Liar dan Taman Negara. *ISSN:01218126 .* (Annual. Vol. 1, 1982 - ). Print

178. **Journal Perak Planters Association.**
Perak Planters Associations. *ISSN:Z090870654.* (1974-1990-). Ceased. Print.

179. **Jurnal al-Tamaddun.**
Universiti Malaya. Jabatan Sejarah dan Tamadun Islam. *ISSN:1823-7517* (Annual. Bil.1, 2005). Hybrid. http://umrefjournal.um.edu.my/public/browse-journal-view.php?id=67.

180. **Jurnal Alam Bina**
Universiti Teknologi Malaysia, Fakulti Alam Bina. *ISSN:15111369 .*(Biannual. Jil.1, 1998 -). Hybrid. http://www.fab.utm.my/Publication-Journal.html.

181. **Jurnal Antarabangsa Teknologi Maklumat**
Universiti Kebangsaan Malaysia. Fakulti Teknologi dan Sains Maklumat. *ISBN:*

Eversion from 2010 issues.
http://plm.org.my/wrdp1/?page_id=322

227. **Jurnal Mekanikal.**
Universiti Teknologi Malaysia. Faculty of Mechanical Engineering. *ISSN:0127-3396* . (Biannual. Vol.1, no.1, 1996 -) Hybrid. Online version from Vol,26, 2008.
http://jurnalmekanikal.fkm.utm.my/?id=Issue&pid=703).

228. **Jurnal Melayu.**
Universiti Kebangsaan Malaysia. Pusat Pengajian Bahasa, Kesusasteraan dan Kebudayaan Melayu. *ISSN:1675-7513.* (Annual. Vol.1, 2004 - ). Hybrid.
http://www.ukm.my/e-melayu/index.php?option=com_content&view=frontpage&Itemid=11&lang=en

229. **Jurnal Paradigma.**
Institut Perguruan Tuanku Bainun. Jabatan Penyelidikan dan Pembangunan Profesionalisme Perguruan (R & D). *ISSN:1985-1731.* (Annual. Jil. 5, 2005 -). Print.

230. **Jurnal Pembacaan**
Persatuan Pembacaan Malaysia. *ISSN:01286188* (Annual, Vol.1 No.1 1992-Vol.2, 2001). Print. Ceased.

231. **Jurnal Pembangunan Sosial.**
Universiti Utara Malaysia. Sekolah Pembangunan Sosial. *ISSN:1394-6528.* (Biannual. Jil.1, 1999 -). Print.

232. **Jurnal Pembangunan Sumber Manusia.**

Pertubuhan Pembangunan Penduduk Malaysia. PP11173-(7). (Annual, Jil.1, 2006-). Print.

233. **Jurnal Pemikir Pendidikan.**
Universiti Malaysia Sabah. *ISSN:1985-3637.* (Biannual, Vol.1 2008-). Print. *Infor. from:*
*http://www.ums.edu.my/webv3/appl/files/pemikir.jpg*

234. **Jurnal Pendidik dan Pendidikan.**
Universiti Sains Malaysia. Pusat Pengajian Ilmu Pendidikan . *ISSN:0126-7272* (Annual. Jil.1, 1979-Jil.18, 2003). Change title: *Asia Pacific Journal of Educators and Education* from 2010). Print.

235. **Jurnal Pendidikan.**
University of Malaya. Faculty of Education. *ISSN:01265261* . (Annual. Jil.1, 1970- Jil.29, 2009). Print. Indexation: MyAIS.

236. **Jurnal Pendidikan Bahasa Melayu.**
Universiti Kebangsaan Malaysia. Fakulti Pendidikan. *ISSN: 2180-4842.* (Biannual, Vol.1, no.1, 2011-). Online.
http://www.ukm.my/jpbm/Current.aspx

237. **Jurnal Pendidikan Guru.**
Kementerian Pelajaran. Bahagian Pendidikan Guru. *ISSN:Z090910397 / 01277316* .(Annual. Jil.1, 1985- Jil. 14, 2001-). Print. Ceased?

238. **Jurnal Pendidikan Malaysia (Malaysian Journal of Education).**
Universiti Kebangsaan Malaysia. Jabatan Pendidikan. *ISSN:0126-6020 /2180-0782.* (Annual and biannual. Vol.1, 1975 -). Hybrid from 2005.
http://www.ukm.my/jurfpend/indexbm.html.Indexation: MyAIS.

239. **Jurnal Pendidikan Matematik dan Sains.**

Kementerian Pendidikan Malaysia . *ISSN:1394-1887.* (Annual, Jil.1 , 1994). Print. Ceased.

240. **Jurnal Pendidikan Teknikal .**
Kementerian Pendidikan Malaysia. Jabatan Pendidikan Teknikal. *ISSN:1675-1604.* (Biannual, Jil.1 Bil.1, 2001-). Print.

241. **Jurnal Pendidikan UPSI.**
Universiti Pendidikan Sultan Idris. Faculty of Education and Human Development. (Annual.Vol.1, 2011). Print.

242. **Jurnal Pendidikan (UTM).**
Universiti Teknologi Malaysia. Jabatan Pendidikan. (Annual. Jil.1,1995-Jil.14, 2009-). Print.

243. **Jurnal Pengajaran dan Pembelajaran Bahasa.**
Kementerian Pelajaran Malaysia. Bahagian Sekolah Sekolah. *ISSN:Z090910400* (Biannual. Jil.1, bil.1, 1984-). Print. Ceased.

244. **Jurnal Pengajian Cina Malaysia**
Huazi Resource & Research Centre Bhd., Kuala Lumpur. *ISSN:1151-0044 : 20.00* (Annual. No. 1, 1997- No.10, 2007). Print. Ceased?

245. **Jurnal Pengajian India**
University of Malaya. Faculty of Arts and Social Sciences. Dept. of Indian Studies. *ISSN:01274929* (Annual. Vol.1, 1983 - Vol 9, 2006-). Print. Ceased.

246. **Jurnal Pengajian Media Malaysia (Malaysian Journal of Media Studies).**
University of Malaya. Faculty of Arts and Social Sciences. Dept. of Media Studies. *ISSN:15112284* . (Annual. Vol.1, no.1, 1998). Hybrid. http://jpmm.um.edu.my/

247. **Jurnal Pengajian Melayu**
University of Malaya. Academy of Malay Studies. *ISSN:1823-7622.* (Annual. Vol.1, 1989 - Vol.21, 2010-). Hybrid. Elektronik from 2010.
http://ejum.fsktm.um.edu.my/VolumeListing.aspx?JournalID=20

248. **Jurnal Pengajian Umum (Journal on General Studies).**
Universiti Kebangsaan Malaysia. Pusat Pengajian Umum. *ISSN:1511-8398* (Annual. 2000 – Bil. 6, 2005). Print.

249. **Jurnal Pengurusan (UKM)**
Universiti Kebangsaan Malaysia. Faculty of Business Administration. *ISSN:01272713* (Annual, Vol.1, 1982). Print. Indexation: Scopus. Tables of contents:
http://www.ukm.my/penerbit/jurus.htm

250. **Jurnal Pengurusan Awam.**
Jabatan Perkhidmatan Awam Malaysia. *ISSN:Z090938305.* (Annual. Jilid 1, 2002). Print.

251. **Jurnal Pengurusan dan Kepimpinan Pendidikan.**
Institut Aminuddin Baki. *ISSN:1511-4147.* (Biannual. Vol.1, 1991 - Vol.15, 2005). Print. Ceased?

252. **Jurnal Pengurusan dan Penyelidikan Fatwa**
Institut Pengurusan & Penyelidikan Fatwa Sedunia, Universiti Sains Islam Malaysia. *ISSN:1675-5936* (Annual. Vol. 1, 2007- ). Print.

253. **Jurnal Pengurusan Pendidikan.**
Institut Aminuddin Baki. *ISSN:0128-2832* (Biannual. Vol.1, 1991-Jil.6, 1997). Print. Ceased?

254. **Jurnal Pentadbiran Sosial Malaysia (Malaysian Journal of Social Administration).**
University of Malaya.Faculty of Arts and Social Sciences. *ISSN:1675-3925* . (Annual. Vol.1, 2002 - Vol.3,2004-). Print. Ceased?

255. **Jurnal Penterjemah**
Persatuan Penterjemah Malaysia. (Annual. Bil.1, 1987). Ceased.

256. **Jurnal Penyelidikan dan Pendidikan Kejuruteraan**
Kolej Universiti Kejuruteraan Utara Malaysia. *ISSN:18322981* (Annual. Jil. 1, 2004-). Print.

257. **Jurnal Penyelidikan Islam.**
Jabatan Perdana Menteri Malaysia. Bahagian Hal Ehwal Islam. *ISSN:Z090937406* . (Annual. Bil. 6, 1986-Bil 20, 2007-). Print.

258. **Jurnal Penyelidikan Maktab Perguruan Sultan Abdul Halim**
Maktab Perguruan Sultan Abdul Halim. *ISSN:1823-6367* (Annual. No.1, 2003). Online. http: //www.ipsah.edu.my/index.php?option=com_docman&task=cat_view&gid=43&Itemid=141

259. **Jurnal Penyelidikan Pendidikan**
Kementerian Pendidikan Malaysia. Bahagian Perancangan dan Penyelidikan Dasar Pendidikan *ISSN:15116530* (Annual. Jil.1, 1999) - Jil.9:(2007) – Incomplete. Print.

260. **Jurnal Penyelidikan Pendidikan Guru**
Kementerian Pelajaran. Bahagian Pendidikan Guru . *ISSN:1823-5891.* (Annual, Jil. 1, 2005- Jil. 5, 2010). Print.

261. **Jurnal Peradaban**
University of Malaya. Pusat Dialog Peradaban. *ISSN:1985-6296* (Annual. Vol.1, 2008-). Print. Indexation status: MyAIS.

262. **Jurnal Peradaban Melayu.**
Universiti Pendidikan Sultan Idris. Institut Peradaban Melayu. *ISSN: 1675-4271.* (Vol.1 2003-). Print.

263. **Jurnal Pergigian Universiti Malaya.**
University of Malaya. Fakulti Pergigian. *ISSN:0127-1369.* (Biannual. Vol.1, no.1, 1983 -). Print. Ceased.

264. **Jurnal Perkama.**
Persatuan Kaunselor Malaysia. *ISSN:0127-6301.* (Annual. Bil.1, 1984 - Bil.15, 2009). Print.

265. **Jurnal Perkhidmatan Sebaran Pendidikan.**
Kementerian Pelajaran Malaysia. *ISSN:01265431* (Jil. 1, 1974 - Jil 15 1987- ). Print. Ceased?

266. **Jurnal Persatuan Linguistik Malaysia**
Persatuan Linguistik Malaysia. *ISSN: 0127-7359* . (Annual. Vol.1, 1983-). Print. E-content available from 2010 issues. http://plm.org.my/wrdp1/?page_id=322 .

267. **Jurnal Persatuan Pendidikan Teknik dan Vokasional Malaysia.**
Universiti Teknologi Malaysia. Persatuan Pendidikan Teknik dan Vokasioanal Malaysia. (Biannual, 2007). Print.

268. **Jurnal Persatuan Perubatan Homeopathy Malaysia.**
Persatuan Perubatan Homeopathy Malaysia. *ISSN:1675-3917.* (Annual. 2005 -). Print.

269. **Jurnal Persuratan Melayu.**
Universiti Kebangsaan Malaysia. Jabatan Persuratan Melayu. *ISSN: 1394-3693.* (Annual, Jil.1, bil.1 (1995 - ). Print. Ceased.

270. **Jurnal Perubatan UKM.**
Universiti Kebangsaan Malaysia. *ISSN:0127-1075.* (Annual and Biannual. Vol. 1, no.1, 1979 - Vol. 17, 1995 -). Print. Ceased?

271. **Jurnal Perumahan.**
Jabatan Perumahan Negara Malaysia *ISSN:0128-5645* (Annual. 1993 - ). Irregular. Print.

272. **Jurnal PPM.**
Persatuan Pustakawan Malaysia *ISSN:1823-6308.* (Annual. Vol.1, 2005 - Vol. 2, 2008-). Print.

273. **Jurnal Produktiviti.**
Pusat Daya Pengeluaran Negara. *ISSN:0127-8223* (Annual. Jil.1, 1986 - Jil 20 2004-) . Print.

274. **Jurnal Psikologi dan Pembangunan Manusia.**
Universiti Kebangsaan Malaysia. Pusat Pengajian Psikologi dan Pembangunan Manusia. *ISSN:1675-4727.* (Anual, Bil.1. 1985- Bil.17, 2001-). Print.

275. **Jurnal Psikologi Malaysia.**
Universiti Kebangsaan Malaysia. Jabatan Psikologi. *ISSN: 01278029.* (Annual. No.1, 1985 - 1993). Print.

276. **Jurnal Psikologi Perkhidmatan Awam Malaysia.**
Jabatan Perkhidmatan Awam. Malaysia. Bahagian Perkhidmatan Psikologi. *ISSN:1823-6197.* (Biannual, Bil. 1 (2006 - 2009-). Print. Tables of content - http://www.interactive1.jpa.gov.my/psikologi/page.php?40.

277. **Jurnal Pusat Sumber Pendidikan Negeri Sabah.**
Pusat Sumber Pendidikan Negeri Sabah. *ISSN:1675-624X.* (Annual. 2004-). Print.

278. **Jurnal Sains.**
University of Malaya. Foundation of Science Centre. *ISSN:0128-7826.* (Annual. Vol.1, 1992 - Vol.15, 2007-). Print. Ceased.

279. **Jurnal Sains dan Matematik.**
Universiti Pendidikan Sultan Idris. *ISSN:19857918.* (Biannual, Vol.1, no.1, 2009) Hybrid. http: //penerbit.upsi.edu.my/website_e-jurnal/jurnal%20site/jsm1.1/content.html.

280. **Jurnal Sains danTeknologi.**
Universiti Tun Hussein Malaysia. *ISSN:1823-2698.* (Annual. Vol.1, no.1, 2004 -). Print.

281. **Jurnal Sains Farmasi Malaysia .**
Malaysian Pharmaceutical Society by Universiti Sains Malaysia. *ISSN:0126-8376.* (Annual. Vol.1, 1978-2002). Changed name to *Malaysian Journal of Pharmaceutical Sciences* in 2003. Printed.

282. **Jurnal Sains Kemanusiaan.**
Universiti Tun Hussein Malaysia.

*ISSN:1675-7580* (Annual. Vol.1, no. 1, 2003 -). Print.

283. **Jurnal Sains Kesihatan Malaysia (Malaysian Journal of Health Sciences).**
Penerbit Universiti Kebangsaan Malaysia. *ISSN:1675-8161* (Annual, No.1, 2003-). Print. Some issues online at http: //www.fsk.ukm.my/jurnal/index.htm

284. **Jurnal Sains Nuklear Malaysia.**
Pusat Penyelidikan Atom Tun Ismail. *ISSN:0127-2810.* (Annual. Jil.1, bil. 1, 1983- Jil. 21, 2009). Print.

285. **Jurnal Sains Sosial.**
Universiti Tun Hussein Malaysia *ISSN:1823-2671* (Annual. Vol.1, no.1, 2003 -). Print.

286. **Jurnal Sejarah Melaka.**
Persatuan Sejarah Malaysia, Cawangan Melaka. (Annual, No.1, 1976 - No. 18 1983). Ceased. Print.

287. **Jurnal Skrin Malaysia.**
Universiti Teknologi MARA. *ISSN:1823-1020* .(Biannual. Jil.1, 2004) Hybrid. http://rmi.uitm.edu.my/jurnal-skrin-malaysia.html)

288. **Jurnal Syariah.**
University of Malaya. Academy of Islamic Studies. *ISSN:01286730.* (Biannual, Vol.1, 1993) . E-version 2002-2008 in MyAIS, http://myais.fsktm.um.edu.my. Indexation status: MyAIS, Index Islamicus.

289. **Jurnal Tanah.**
Persatuan Pengurusan Tanah Semenanjung Malaysia. *ISSN:1675-3046* (Semiannual,Vol.1, 1992-). Print.

290. **Jurnal Teknologi.**
Universiti Teknologi Malaysia. *ISSN:0127-9696.* (Annual. Bil 18, 1991). Hybrid. Indexation: MyAIS. http://www.penerbit.utm.my/cgi-bin/jurnal/jurnal.cgi.

291. **Jurnal Teknologi Maklumat.**
Universiti Teknologi Malaysia. Institut Sains Komputer. *ISSN:0128-3790* (Biannual. Jil. 1, 1990- Jil.20 , 2008- ). Print.

292. **Jurnal Teknologi Maklumat dan Multimedia.**
Universiti Kebangsaan Malaysia Press. *ISSN:18230113* (Annual. Vol.1, 2004)- Vol. 3, 2006--). Print.

293. **Jurnal Teknologi Maklumat dan Sains Kuantitatif.**
Universiti Teknologi MARA. *ISSN:1823-0822* . (Annual. Vol.1, 1999 –Vol.8, 2006-). Print.

294. **Jurnal Ukur Bahan.**
Universiti Teknologi Malaysia. Fakulti Alam Bina. (Annual. 1996 - ). Print

295. **Jurnal Undang-Undang dan Masyarakat.**
Universiti Kebangsaan Malaysia. Faculty of Law. I*SSN:13947729* . (Annual. Vol. 1, 1997- Vol.11, 2007-). Print.

296. **Jurnal Undang-Undang IKIM.**
Institut Kefahaman Islam Malaysia. *ISSN:1511-0281.* (Biannual. Vol.1, 1997 - ). Print.

297. **Jurnal Usuluddin.**
University of Malaya. Academy of Islamic Studies. *ISSN:1394-3723.* (Annual. No. 1,

1993 – No.29, 2009). Print. Indexation status: MyAIS

298. **Jurnal Veterinar Malaysia.**
Veterinary Association of Malaysia. *ISSN:9128-2506* (Biannual.Vol.1, no.1, 1966-Vol.9, 1997). Print.

299. **Jurnal Warisan Johor.**
Yayasan Warisan Johor. *ISSN:1511-1377.* (Annual. Jil.1, 1997 - Jil.7, 2003). Ceaded? Print.

300. **Jurnal Yadim.**
Yayasan Dakwah Islamiah Malaysia. *ISSN:1511-905X.* (Biannual. Jil.1, 2001-Jil.10, 2008). PNM. Print.

301. **Jurutera Galian: Bulletin of Institute of Minerals Engineering, Malaysia.**
Institute of Mineral Engineering. *ISSN:9090-9402.* (Quarterly. No 5, 1981 - no 22, 1986-). Print. Ceased?

302. **JUTEKS, Jurnal Teknikal & Kajian Sosial.**
Kolej Sains dan Teknologi, Universiti Teknologi Malaysia International Campus. *ISSN:1675-2228.* (Biannual. Jil. 1, 2002- ). Print.

303. **Kajian Ekonomi Malaysia.**
Persatuan Ekonomi Malaysia. (Annual. Vol.1, 1964 - Vol. 24, 1987-). Print.

304. **Kajian Malaysia (Journal of Malaysian Studies).**
Universiti Sains Malaysia. *ISSN:0127-4082* (Biannual. Vol.1, 1983). Hybrid. http://web.usm.my/km/forthcoming.html.

305. **Kajian Veterinar Malaysia.**
Association of Veterinary Surgeons Malaysia. *ISSN:0047-309X* .(Biannual. Vol.1, no.1, 1974- Vol.19, no.2, 1987) Ceased. Print.

306. **Kanun: Jumal Undang-Undang Malaysia.**
Dewan Bahasa & Pustaka. (Quarterly. Vol.1, No.1, 1989 -). Print.

307. **KATHA - The Official Journal of the Centre for Civilisational Dialogue**
University of Malaya, Centre for Civilisational Dialogue. *ISSN:1823-2159.* (Annual. Vol. 1, 2005). Print. Indexation: MyAIS.

308. **Kedah dari Segi Sejarah.**
Persatuan Sejarah Malaysia. *ISSN:Z09080435X* .(Annual. Jil.1, 1966-Jil.8, 1979). Ceased. Print.

309. **Kekal Abadi**.
University of Malaya Library. *ISSN:0127-2578.* (Quarterly and later annual. Jil.1, bil.1,1982 -). Hybrid. http://www.umlib.um.edu.my/scontent s.asp?tid=31&cid=100&p=1&vs=?vs=en. Indexation status: MyAIS.

310. **Kemanusiaan: the Asian Journal of Humanities.**
Universiti Sains Malaysia. Pusat Pengajian Ilmu Kemanusiaan. *ISSN:1394-9330* . (Annual. 2008-) Former title: *Jurnal Ilmu Kemanusiaan.* Hybrid. http://web.usm.my/kajh/. Indexation: MyAis, DOAJ.

311. **Kesturi: jurnal Akademi Sains Islam Malaysia.**
Akademi Sains Islam Malaysia. *ISSN:15113078* . (Biannual. Vol.1, no.1, 1991- Jil.11, 2001 ). Ceased. Print.

312. **Kinabalu: Jurnal Perniagaan & Sains Sosial.**

Universiti Malaysia Sabah.*ISSN:1394-4517* .(Annual. Bil.1, 1995 - Bil.11, 2005). USM. Ceased? Print.

313. **Labuan Bulletin of International Business & Finance**.
Universiti Malaysia Sabah. Labuan School of International Business & Finance. *ISSN: ISSN 1675-7262*. (Biannual. Vol.1, no.1, 2003 -). Hybrid. http://wwwkal.ums.edu.my/lbibf/. Indexation status: MyAIS, Google Scholar

314. **Labuan e-Journal of Muamalat and Society**
Universiti Sabah Malaysia *ISSN:1985-482X* (Annual.Vol.1, 2007). Online. http://wwwkal.ums.edu.my/ljms/. Indexation status: MyAIS.

315. **Majalah Perpustakaan Malaysia**
Persatuan Perpustakaan Malaysia. *ISSN:0126-7809* .(Annual. Vol.1, 1972 – Vol.14, 1989). Ceased. Print.

316. **Malawati : Jurnal Sejarah Selangor Darul Ehsan.**
Persatuan Sejarah Malaysia. Cawangan Selangor. *ISSN:0127-953X.* (Annual. 1996 -). Ceased. Print.

317. **Malay literature (DBP).**
Dewan Bahasa dan Pustaka. *ISSN:0128-1186* . (Annual. Vol.1, no.1, 1988-).Print. Website: http://malayliterature.wordpress.com/about/.

318. **Malayan Agriculturist.**
University of Malaya. Agriculture Society. *ISSN:Z001265407* . (Annual. Vol. 1, 1960/61- Vol.13, 1974-75-). Ceased. Print.

319. **Malayan Economic Review.**
Malayan Economics Society. (Biannual. Vol 1, 1956 - Vol.27, 1982). Ceased. Print.

320. **Malayan Forester.**
Federation of Malaya. Officers of the Forest Department. ISSN: *Z00025-1275.* (Annual. Vol.1, 1931/31- Vol.35, 1972-). Ceased. Print.

321. **Malayan Historical Journal.**
Malayan Historical Journal. *ISSN:9081-1275.* (Annual. 1954-1956). Ceased. Print.

322. **Malayan Journal of Tropical Geography.**
University of Malaya. Dept of Geography. *ISSN:Z090839765.* (Vol. 1, 1953 - Vol.10, 1957). Ceased. Print.

323. **Malayan Law Journal.**
Malaya Publishing House. *ISSN:0025-1283* . (Monthly. Vol. 1, 1932-). Print.

324. **Malayan Library Journal.**
Persatuan Perpustakaan Malaysia. (Quarterly. Vol.1, 1960 - Vol.3,1964). Ceased. Print.

325. **Malayan Nature Journal.**
Malayan Nature Society. *ISSN:0025-1291.* (Quarterly. Vol.1, 1940/1941-). Print. Indexation status: Scopus.

326. **Malayan Scientist.**
University of Malaya. Science Society. *ISSN:Z090819098* (Vol. 1, 1963/64-Vol.6, 1971/72-). Ceased. Print.

327. **Malaysia dari Segi Sejarah.**
Persatuan Sejarah Malaysia.

*ISSN:9081-150X* (Annual.Vol.1, 1962 - Vol.36, 2008-). Print.

328. **Malaysia in History.**
Malaysian Historical Society. *ISSN:0047-5610* . (Annual. Vol. 1, 1954 - Vol.28, 1985-). Ceased? Print.

329. **Malaysian Accountant**
Malaysian Association of Certified Public Accountants. *ISSN:01270087.* (Quarterly. Vol.1, 1980-). Print.

330. **Malaysian Accounting Review**
Universiti Teknologi MARA and Malaysian Accountancy Research & Education Foundation . *ISSN:1675-4077.* (Annual. Vol.1, 2002-). Hybrid. http://www.maref.org.my/publication/malaysian-accounting-reviews.html.

331. **Malaysian Agricultural Journal.**
Kementerian Pertanian Malaysia. (Annual. 1935 - 1993). Ceased. Print.

332. **Malaysian Applied Biology Journal. (Jurnal Biologi Gunaan Malaysia).**
Malaysian Society of Applied Biology. *ISSN:0126-8643.* (Biannual. Vol.1, 1977). Hybrid. http://www.mabjournal.com/.

333. **Malaysian Dental Journal.**
Malaysian Dental Association. *ISSN:Z090872487* . (Biannual. Vol.1, 1961-Vol.31, 2010). Print.

334. **Malaysian Family Physician**
Academy of Family Physicians of Malaysia. *ISSN:1985-2274.* (Triannual. 1989-2004-print. Hybrid. E-version from Vol.13 no.3, 2005). Hybrid. http://www.e-mfp.org/. Indexation: Scopus.

335. **Malaysian Fisheries Journal**
Fisheries Research Institute, Department of Fisheries Malaysia *ISSN:1511-7286* (Annual.Vol.1, 2002) - Vol.6 (2007-). Print.

336. **Malaysian Forester.**
Institut Penyelidikan Perhutanan Malaysia. *ISSN:0302-2935* .(Biannual. Vol. 36, 1973 -1983 ) Print. Indexation: Scopus-2007-.

337. **Malaysian Geographers.**
National Geographical Association of Malaysia. *ISSN:0126-624X* . (Annual. Vol.1, 1978 - Vol.3, 1981). Ceased. Print.

338. **Malaysian Journal of Agricultural Economics.**
Universiti Pertanian Malaysia. *ISSN:9090-0537.* (Vol.1, 1984 - Vol 12, 1998-). Print.

339. **Malaysian Journal of Analytical Sciences.**
Persatuan Sains Analisis Malaysia . *ISSN:1394-2506* (Biannual. Vol.1,No.1&2:1995-.) Hybrid. http://www.ukm.my/mjas/) Indexation status: Scopus; MyAIS

340. **Malaysian Journal of Biochemistry and Molecular Biology.**
Malaysian Biochemical Society. *ISSN:1511-2616.*(Annual. Vol.11, 2005 - Vol.17, 2009-). Hybrid. http://ejum.fsktm.um.edu.my/VolumeListing.aspx?JournalID=16. Indexation: MyAIS.

341. **Malaysian Journal of Chemistry.**
Malaysian Institute of Chemistry. (Annual. Vol.1,1999-). Online. http://www.ikm.org.my/index.php?option=com_content&view=article&id=97&Itemid=114.

342. **Malaysian Journal of Child Health.**
Malaysian Paediatric Association. *ISSN:1394-0090.* (Annual. Vol.1, 1988 - Vol.6, 1994-). Ceased. Print.

343. **Malaysian Journal of Civil Engineering**.
Universiti Teknologi Malaysia. Faculty of Civil Engineering. ISSN: 1823-7843.(Biannual. Vol.1, 1988?-). E-version from Vol.7, 1994. Hybrid. http://www.utm.my/civil/content.php?id=389&cid=204&lang=1. Indexation status: MyAIS

344. **Malaysian Journal of Community Health Malaysia.**
Universiti Kebangsaan Malaysia Medical Centre. *ISSN: 1675-1663.* (Biannual, Vol.1, 1994-). Previous title: *Journal of Community Health, Jurnal Kesihatan Masyarakat.* Hybrid. http://www.communityhealthjournal.org/. Indexation status: Index Medicus, Medline

345. **Malaysian Journal of Computer Science.**
University of Malaya. Faculty of Computer Science & Information Technology. *ISSN:01279084* (Biannual. 1987-).E- version, 1996-. Hybrid. http://ejum.fsktm.um.edu.my/VolumeListing.aspx?JournalID=4. Indexation status: SCI, Scopus, Inspec.

346. **Malaysian Journal of Consumer and Family Economics.**
Malaysian Consumer and Family Economics Association. *ISSN:1511-2802* . (Annual. Vol.1, 1998-). Hybrid. http://bakkdev.um.edu.my/myjurnal/public/browse-journal-view.php?id=108. Indexation: Scopus.

347. **Malaysian Journal of Dermatology (Jurnal Dermatologi Malaysia).**
Persatuan Dermatologi Malaysia. *ISSN:1511-5356* . (Annual. Vol. 1, 1988 -). Mainly print. 2007-2009 available at: http://www.dermatology.org.my/journal.htm

348. **Malaysian Journal of Economic Studies.**
University of Malaya. Faculty of Economics and Administration. Former title: *Kajian Ekonomi Malaysia* .*ISSN:9093-9751.* (Biannual. Vol 25, 1988 - Vol.48, 2011-). Print. Table of contents:http://www.pem.org.my/mjes.html. Indexation: Scopus, ABI/Inform, EconLit.

349. **Malaysian Journal of Educational Technology**
Malaysian Educational Technology Association. *ISSN:1675 0292.* (Annual. Vol.1, no.1, 2001-). Print. Table of contents: http://www.myjet-meta.com/past-issues.php.

350. **Malaysian Journal of ELT Research.**
Malaysian English Language Teachers Association. *ISSN:15118002.* (Annual, Vol.1, 2005-). Hybrid. http://www.melta.org.my/home/?domain=melta&doit=showclass&cid=2).

351. **Malaysian Journal of Family Studies**
National Population and Family Development Board. *ISSN:0128-1232* (Annual. No.,1989 - No 2, 1990). Ceased. Print.

387. **Malaysian Orthopaedic Journal**
Malaysian Orthopaedic association
*ISSN:1985-2533.*(Biannual. Vol.1 no.1, 2007-). Hybrid.
http://www.morthoj.org/about_moj.html. Indexation status: MyAIS.

388. **Malaysian Polymer Journal**.
Universiti Teknologi Malaysia. Faculty of Chemical and Natural Resources Engineering. ISSN : 1823-7789. (Biannual. Vol.1, no.1, 2006-). Online: http://www.cheme.utm.my/mpj/index.php?option=com_content&task=view&id=15&Itemid=29. Indexation status: MyAIS.

389. **Malaysian Tax Journal**
Inland Revenue Officer's Union Peninsular Malaysia.*ISSN:0126-5075.* (Annual. Vol.1, 1974- Vol.20, 1990). Ceased? Print.

390. **Malaysian Technologist**
Technological Association of Malaysia. *ISSN:Z090894529.* (Annual. Vol.1, 1984 - 1996-). Ceased. Print.

391. **Malim: Jurnal Pengajian Umum Asia (Southeast Journal of General Studies).**
Universiti Kebangsaan Malaysia. Pusat Pengajian Umum. (Annual. Vol.1, 2000-2009). Hybrid from 2005.
http://www.ukm.my/jmalim/.Also know as *Jurnal Pengajian Umum Asia.*

392. **Manu : Jurnal Pusat Penataran Ilmu & Bahasa.**
Universiti Malaysia Sabah. Pusat Penataran Ilmu & Bahasa. *ISSN:1511-1989.* (Annual, Bil.1, 1998-). Print.

393. **Manusia dan Masyarakat : Siri Baru**
University of Malaya. Faculty of Arts and Social Sciences, Dept of Sociology & Anthropology . *ISSN: 9083-0326*(Annual. Vol.1, 1978 - Vol.19, 2010-). Print.

394. **MARDI Research Journal.**
MARDI ISSN: 0128-0686. (Annual. Vol. 16, 1988 – Vo. 24, 1996). Ceased. Print.

395. **Masa : Jurnal Pusat Penyelidikan Islam.**
Pusat Penyelidikan Islam Malaysia, Jabatan Perdana Menteri. (Annual. Th.1, bil.1, 1978 -1985). Ceased. Print.

396. **Masalah Pendidikan**
University of Malaya. Faculty of Education. *ISSN:0126-5024* (Annual. Jil.1, 1965 - Jil. 31, 2008-). Print. Indexation status: MyAIS.

397. **Matematika Jurnal**
Universiti Teknologi Malaysia. Faculty of Science. *ISSN:0127-8274.* (Annual. Vol.1, 1985 - Vol. 24, 2008-). Hybrid. http://www.fs.utm.my/matematika/content/section/4/31/. Indexation status: MyAIS.

398. **Medical & Health Reviews**
Universiti Teknologi MARA. *ISSN:1985-6512.* (Annual. Vol.1 no.1, 2008-). Hybrid. http://mhr.uitm.edu.my/.

399. **Medical Journal of Malaysia**
Medical Association of Malaysia *ISSN:0300-5283* (Quarterly. Vol. 27, 1972 - Vol. 63, 2008-). Hybrid. http://www.e-mjm.org/past_issues.html. Indexation status: Scopus, Medline, MyAIS.

400. **Medicine and Health: Official Journal of the Faculty of Medicine, UKM.**
Universiti Kebangsaan Malaysia. Faculty of Medicine. *ISSN:1823-2140 .* (Biannual. Vol.1,no.1,2006). Hybrid. http://www.ppukm.ukm.my/ukmmcjournal/.

401. **Melayu: Jurnal Antarabangsa Dunia Melayu.**
Dewan Bahasa dan Pustaka. *ISSN:1675-6460.* (Annual, Jil.1, 2003 - Jil.5, 2006-). Print.

402. **Menemui Matematik.**
Malaysian Mathematical Society. *ISSN:01269003.* (Biannual. Vol.1, 1979- Vol.27, 2005-). Print.

403. **MII Journal.**
Malaysian Insurance Institute. *ISSN:9093-6000* .(Annual. Vol.1, 1986 - Vol 7, 1992). Ceased. Print.

404. **MOST: Malaysian Oil Science and Technology.**
Malaysian Oil Scientists' and Technologists Association. (Biannual. Vol.1, 1990-). Print.

405. **Muaddib: Jurnal Penyelidikan Maktab Perguruan Kuala Terengganu.**
Maktab Perguruan Kuala Terengganu (Annual. Vol.1, 2001-2005). Print.

406. **Nature Malaysiana.**
Tropical Press . *ISSN: 0126-5318.* (Annual.Vol.1, 1976 - Vol.22, 1997-). Print.

407. **NST Quarterly.**
Malaysian Institute for Nuclear Technology Research. *ISSN:1394-4878.* (Quarterly. 1993 -). Print.

408. **Oil Palm Industry Economic Journal.**
Lembaga Minyak Kelapa Sawit Malaysia (MPOB). *ISSN:1675-0632 .* (Biannual, Vol.1 No.1, 2001 - ). Hybrid. http://palmoilis.mpob.gov.my/publications/opiej.html.

409. **Options (UPM).**
Universiti Pertanian Malaysia. (Annual. Vol.1,1986-Vol.11,2006-). Hybrid. http://ikdpm.upm.edu.my/www/bm/penerbitan/options

410. **Pemimpin: Jurnal Institut Pengetua**
University of Malaya. Institute Pengetua. (Annual. Vol.1, no.1, 2001 - V.8, 2008-). Print.

411. **Pendeta : Jurnal Mengajar Bahasa dan Sastera.**
Universiti Pendidikan Sultan Idris. *ISSN:1823-6812.* (Biannual. Vol.1, no.1, 2006- ). Print.

412. **Penulis Jurnal Persatuan Penulis Nasional Malaysia.**
Persatuan Penulis Nasional. *ISSN :0127-2377.* (Annual. Vol.1, 1964-). Print.

413. **Pertanika.**
Universiti Pertanian Malaysia. (Biannual, Vol.1 no.1, 1978 - Vol.15, 1992-). Ceased. Print.

414. **Pertanika: Journal of Tropical Agricultural Science.**
Universiti Pertanian Malaysia. *ISSN:1511-3701.* (Biannual. Vol.1, no.1, 1978 -). Hybrid. Indexation: Scopus. http://www.pertanika.upm.edu.my/index%20%20Archives%20(PERTANIKA).htm.

415. **Pertanika: Science & Technology.**
Universiti Pertanian Malaysia. *ISSN: 0128-7680 .* (Annual. Vol.1, 1993-). Hybrid. Indexation: Scopus. http://www.pertanika.upm.edu.my/index%20-%20Archives%20(JST).htm

416. **Pertanika: Social Sciences & Humanities.**
Universiti Pertanian Malaysia. *ISSN:0128-7702 .* (Biannual: Vol.1, 1993). Hybrid. Indexation: Scopus. http://www.pertanika.upm.edu.my/index%20-%20Archives%20(JSSH).htm.

417. **Planning Malaysia: Journal of the Malaysian Institute of Planners**
Malaysian Institute of Planners *ISSN:1675-6215* (Annual. Vol.1, 2003-Vol-7, 2009-). Print.

418. **Porim Annual Research Review**
Institiut Penyelidikan Minyak Kelapa Sawit Malaysia. *ISSN:1364-0333 .* (Annual. Vol.1, 1995 - 2001-). Print.

419. **Public Sector ICT Management Review.**
Institut Tadbir Awam Malaysia. *ISSN: 1985-0778.* (Annual. Vol.1, no.1, 2007). Hybrid.http://apps.intan.my/psimr/vol1.1a.html.

420. **Purba: Jurnal Persatuan Muzium Malaysia.**
Persatuan Muzium Malaysia. *ISSN: 0127-1857.* (Annual. No.1, 1982- No.25, 2006). Print.

421. **Rumpun Melayu-Polinesia.**
Universiti Kebangsaan Malaysia *ISSN:0128-3146* (Biannual. Bil. 1, 1990 - Bil.17, 2004). Print. Contents: http://www.ukm.my/patma/rumpun.html.

422. **Sabah Museum and Archive Journal.**
Sabah Museum and State Archives Dept (Annual. No.1, 1986 - No.3, 1990). Ceased. Print.

423. **Sabah Museum Journal.**
Department of Sabah Museum *ISSN:0128-7095* (Annual. Vol.1, no.2, 1994-). Print.

424. **Sabah Society Journal.**
Sabah Society, Kota Kinabalu. *ISSN:0036-2131.*(Annual. Vol.1, 1961 - Vol.20, 2003-). Print.

425. **Sains Malaysiana.**
Universiti Kebangsaan Malaysia. Faculty of Science and Technology. *ISSN:0126603.* (Quarterly. Vol.1, no.1, 1972 -). Hybrid. http://www.ukm.my/jsm/contents.html . Indexation status: ISI (SCI), JCR, Scopus, Chemical Abstracts, Zentralblatt MATH, Zoological Record, MyAIS.

426. **Sandakania.**
Forest Research Center. Forestry Dept. Sabah, Malaysia. *ISSN:01285939.* (Biannual. Vol.1, 1992 - No.17, 2008-). Print.

427. **Sarawak Museum Journal.**
Sarawak Museum. *ISSN:0375-3050.* (Biannual.No.74, 1998-No.87,2010-). Print.

428. **SARE (Southeast Asian Review of English).**
Dept of English, Faculty of Arts and Social Sciences, University of Malaya for the association for Commonwealth Literature & Language Studies in Malaysia.*ISSN:0127-046X* (Annual. Vol.1, 1980 - Vol.47 (2006/07-). Print.

429. **Sari : Jurnal Alam dan Tamadun Melayu (ATMA)**
Universiti Kebangsaan Malaysia. Institut Bahasa, Kesusasteraan dan Kebudayaan Melayu.*ISSN:0127-2721* (Annual, Vol.1, 1987- Vol.27, 2009-).Hybrid.